\documentclass[fleqn,10pt]{wlscirep}
\usepackage[utf8]{inputenc}
\usepackage[T1]{fontenc}

\usepackage{graphicx}
\usepackage{subfigure}
\usepackage{caption}
\usepackage{tikz}
\usetikzlibrary{spy}
\usepackage{siunitx}

\DeclareMathOperator{\Fop}{\mathcal{F}}

\DeclareMathOperator{\Gop}{\mathcal{G}}

\DeclareMathOperator{\Qop}{\mathcal{Q}}

\newcommand\norm[1]{\left\lVert#1\right\rVert}

\usepackage{algpseudocode}

\usepackage{textcomp}

\usepackage[noend,ruled,linesnumbered]{algorithm2e}
\SetAlFnt{\small}
\SetAlCapFnt{\normalsize}
\SetAlCapNameFnt{\normalsize}
\SetKwProg{Pro}{Procedure}{}{}
\SetKwProg{Fn}{Function}{}{}
\SetKwBlock{Ifd}{\#ifdef}{\#endif}
\SetKwProg{With}{with}{ }{}
\SetKwProg{Withdo}{with}{ do}{}
\SetKwRepeat{Dowhile}{do}{while}
\makeatletter

\newcommand{\nosemic}{\renewcommand{\@endalgocfline}{\relax}}
\newcommand{\dosemic}{\renewcommand{\@endalgocfline}{\algocf@endline}}

\makeatother

\SetCommentSty{xCommentSty}

\let\oldnl\nl
\newcommand{\nonl}{\renewcommand{\nl}{\let\nl\oldnl}}

\newif\iffinal

\iffinal
    \newcommand\xiaodong[1]{}
\else 
    \newcommand\xiaodong[1]{{\color{red}[Xiaodong: #1]}}
\fi

\title{Scalable and accurate multi-GPU based image reconstruction of large-scale ptychography data}

\author[1,*]{Xiaodong Yu}
\author[2]{Viktor Nikitin}
\author[2]{Daniel J. Ching}
\author[2]{Selin Aslan}
\author[2,3]{Do\u ga G\" ursoy}
\author[1,2,*]{Tekin Bi\c cer}
\affil[1]{Data Science and Learning Division, Argonne National Laboratory, 9700 Cass Avenue, Lemont, Illinois
60439, USA}
\affil[2]{X-ray Science Division, Argonne National Laboratory, 9700 Cass Avenue, Lemont, Illinois 60439, USA}
\affil[3]{Department of Electrical Engineering and Computer Science, Northwestern University, 2145 Sheridan Road, Evanston, Illinois 60208, USA}

\affil[*]{\{xyu, tbicer\}@anl.gov}

\begin{abstract}
While the advances in synchrotron light sources, together with the development of focusing optics and detectors, 
allow nanoscale ptychographic imaging of materials and biological specimens, 
the corresponding experiments can yield terabyte-scale large volumes of data that can impose a heavy burden on the computing platform.
While Graphical Processing Units (GPUs) provide high performance for such large-scale ptychography datasets, a single GPU is typically insufficient for analysis and reconstruction.
Several existing works have considered leveraging multiple GPUs to accelerate the ptychographic reconstruction. 
However, they utilize only Message Passing Interface (MPI) to handle the communications between GPUs. 
It poses inefficiency for the configuration that has multiple GPUs in a single node, especially while processing a single large projection, since it provide no optimizations to handle the heterogeneous GPU interconnections containing both low-speed links, e.g., PCIe, and high-speed links, e.g., NVLink.
In this paper, we provide a multi-GPU implementation that can effectively solve large-scale ptychographic reconstruction problem with optimized performance on intra-node multi-GPU.
We focus on the conventional maximum-likelihood reconstruction problem using conjugate-gradient (CG) for the solution and propose a novel hybrid parallelization model 
to address the performance bottlenecks in CG solver. 
Accordingly, we develop a tool called PtyGer (\textbf{Pty}chographic \textbf{G}PU(multipl\textbf{e})-based
\textbf{r}econstruction), implementing our hybrid parallelization model design.
The comprehensive evaluation verifies that PtyGer can fully preserve the original algorithm's accuracy while achieving 
outstanding intra-node GPU scalability.
\end{abstract}
\begin{document}

\flushbottom
\maketitle
%
%
\thispagestyle{empty}


\section*{Introduction}\label{sec:intro}

Coherent diffraction imaging (CDI) is a lensless technique that can produce nanometer-scale resolution images by avoiding lens-imposed limitations of the traditional microscopy~\cite{Abbey:08, Dierolf:08, Chapman:10, Nugent:10, Miao:12, Miao:15}.
Ptychography~\cite{Hoppe:69} is a scanning CDI technique that has gained popularity due to the increasing brilliance and coherence of the synchrotron light sources~\cite{pfeiffer2018x}. 
Ptychography benefits from the advantages of both CDI and the scanning-probe microscopy by using a focused beam of light (a.k.a the probe) to scan the object at a series of overlapping scan positions, and collecting the corresponding diffraction patterns at the far-field by a pixelated detector. 
It can be extended to 3D by rotating the object to different view angles and repeating the scanning process at each view to yield a tomographic dataset~\cite{aslan2019joint} or by using the new acquisition methods that are based on rotation as the “fast” scan-axis~\cite{ching2018rotation}.
A range of image reconstruction models and iterative solution methods~\cite{Faulkner:04,Rodenburg:04,Maiden:09,Thibault:08,Thibault:09,guizar2008phase,Thibault:12,odstrvcil2018iterative} 
has been proposed to reconstruct the object based on the solutions of the phase-retrieval problem. 
Because the data volume is high that can reach to terabytes, all these approaches are compute-intensive and can require a significant amount of time when running on a single processing unit, either CPU or GPU~\cite{dong2018high}.

Over the last decade, Graphics Processing Units (GPUs) have been broadly used for general-purpose computing due to their massive parallelism and computational power. 
The applications from various domains have been successfully accelerated with GPUs, including network intrusion detection~\cite{Yu_ppopp,Yu:2013:GAR:2482767.2482791,yu_ics,yu_thesis}, 
biological sequence alignment~\cite{hou2016aalign,zhang2017cublastp}, program analysis~\cite{yu2020ipdps,yu2019dissertation}, 
and tomographic reconstruction~\cite{yu_ccgrid,gursoy2014tomopy,Yu_cf17,hidayetouglu2019memxct,yu2019GPU}, to name a few.
Several existing works~\cite{nikitin2019photon,Wakonig:zy5001} have implemented the standard algorithms for ptychographic reconstruction on a single GPU. 
However, still the memory of a GPU falls in short compared to the data volumes of even a single view produced by ptychography experiments (see Sec. Reconstruction efficiency).
Furthermore, the upcoming upgrade of the light source facilities will provide up to two orders of magnitude improvement in beam brightness, which will lead to a proportional increase in ptychography dataset sizes.  
These challenges motivate us to leverage multiple GPU devices, and therefore utilize aggregated memory, to parallelize and accelerate ptychographic reconstruction.

Most of the current state-of-the-art high-performance computing systems are configured as multi-node multi-GPU
machines\cite{hines2018stepping,papka20182018,gayatri2019comparing}. Accelerating applications on these machines demand 
proper synchronizations and communications among the GPU devices. 
A few existing works~\cite{nashed2014parallel,marchesini2016sharp,dong2018high} 
implement ptychography on multi-GPU with various 
reconstruction algorithms including ePIE~\cite{Maiden:09}, relaxed averaged alternating 
reflections (RAAR)~\cite{luke2004relaxed}, and difference map (DM)~\cite{Thibault:08}. 
However, they all utilize only Message Passing Interface (MPI) to handle the communications. 
MPI~\cite{thakur2005optimization}, the \textit{de facto} standard of inter-node 
communication handling, has been well established, and its functionality has been extended to be GPU-aware~\cite{wang2013gpu,awan2017s,awan2019scalable}.
However, it has sub-optimal performance for intra-node GPU-GPU communications due to the factors such as the heavy 
overhead for shared memory model and the under-utilization of the high-speed low-speed mixed links.
Recently, NVIDIA has proposed NVIDIA Collective Communications Library (NCCL)~\cite{NCCL} for multi-GPU communications. It provides lightweight 
and optimized multi-threading solutions for handling intra-node GPU-GPU communications. 
Coupling NCCL with MPI can offer optimal performances for multi-GPU computing that requires both inter-node 
and intra-node communications.

In this paper, we introduce a parallelization design for solving the maximum-likelihood (ML) ptychographic
reconstruction problem using a conjugate-gradient (CG) solver~\cite{nikitin2019photon} on multi-GPU systems. Our design incorporates 
multi-threading GPU-GPU communications to enables reconstruction of large views with aggregated GPU memories.
CG solver is known to converge faster than the gradient descent based approaches (such as parallel ePIE) and is fully scalable. 
However, implementation of CG on multi-GPU exposes challenges in
preserving the full algorithmic equivalence because of the required scatter-gather operations (see Sec. Challenges of CG solver parallelization). Towards this end, our work proposes 
a novel parallelization model to fit the multi-GPU CG solver. In our model, we split each reconstruction iteration into four stages and then apply different communication patterns 
to different stages to accomplish various data transfer requests. This advanced design maintains the full equivalence of the original CG solver and 
minimizes the multi-GPU implementation overhead. 
Subsequently, we implemented our solution in the \textbf{Pty}chographic
\textbf{G}PU(multipl\textbf{e})-based \textbf{r}econstruction that we call PtyGer, an openly available software package.
Our experiments demonstrate that PtyGer can provide reconstructions with remarkable multi-GPU scalability while maintaining the accuracy of the solution.
In addition, PtyGer can be easily extended to 3D ptychography thanks to the data independence between 
the view angles and the tomo-ptycho joint solvers~\cite{aslan2019joint}. PtyGer also provides interfaces 
for other scalable algorithms that follows a similar data processing pattern~\cite{Thibault:08,Thibault:09,enfedaque2019high} 
to be plugged into.
Our contributions can be summarized as follows:
\begin{itemize}
    \item We propose a fine-grained parallelization design for ptychographic reconstruction on multiple GPUs.
    We explore the challenges of parallelizing the conjugate-gradient solver based maximum-likelihood phase-retrieval algorithm, and accordingly provide a novel hybrid model to tackle the challenges.
    \item We develop a ptychographic reconstruction software called PtyGer. PtyGer implements our hybrid model design using Python and CUDA. It is scalable in terms of both dataset volume and GPU configuration.
    \item We extensively evaluate the efficacy and efficiency of PtyGer. We use both synthetic and real-world experimental dataset with various data volumes. The results show that PtyGer can provide accurate reconstruction with outstanding multi-GPU scalability.
\end{itemize}

\section*{Related work}\label{sec:relatedwork}

A variety of ptychographic reconstruction algorithms have been proposed during the past two decades. 
The ptychographic iterative engine (PIE)~\cite{Faulkner:04,Rodenburg:04} is an iterative phase-retrieval approach 
employing the diffraction patterns as a known illumination function. It demonstrates fast convergence rate~\cite{rodenburg2007transmission,rodenburg2007hard}. However, it lacks the robustness when noise is present in measurements.
Moreover, the PIE and its variants (ePIE\cite{Maiden:09,maiden2017further}, 3PIE~\cite{maiden2012ptychographic}) are 
inherently sequential since they have to update probe and object images after processing each diffraction pattern.
In contrast to the PIE family, some reconstruction algorithms are parallelizable~\cite{enders2016computational}, 
including the difference map (DM) algorithm~\cite{Thibault:08,Thibault:09} and the relaxed averaged alternating 
reflections (RAAR) algorithm~\cite{luke2004relaxed}. Another notable parallel-friendly 
ptychographic reconstruction approach uses the maximum-likelihood (ML) model~\cite{Thibault:12,odstrvcil2018iterative} 
and is derived from the cost–function optimization technique~\cite{guizar2008phase}.
The DM and RAAR algorithms and ML-based models that use CG solver are all naturally parallelizable since they only need to update the estimated 
object image after simultaneously processing a set of diffraction patterns. 
Recently, some advanced techniques~\cite{aslan2019joint,deng2019velociprobe} 
have been proposed to further refine the reconstructed images.

There are several existing works that discuss the GPU-based implementations of ptychography. 
PyNX~\cite{mandula2016pynx} implements the DM algorithm on the GPU using OpenCL language. 
PtychoShelves~\cite{Wakonig:zy5001} provides a MATLAB-based GPU reconstruction engine that supports the DM and ML algorithms. 
However, both of them only allow reconstructions on a single GPU. 
Marchesini et al. developed a multi-GPU-based ptychographic solver called SHARP~\cite{marchesini2016sharp} 
implementing the RAAR algorithm. Dong et al. implemented the DM algorithm on multiple GPUs~\cite{dong2018high}. 
But they utilize only MPI to handle both intra- and inter-node multi-GPU processing, and thus are sub-optimal. Although CUDA-aware MPI 
enables direct GPU-GPU data access and transfer, MPI is not optimized for the intra-node multi-GPU 
communications due to inefficient utilization of high-speed links, e.g., NVLink.
PtychoLib~\cite{nashed2014parallel} proposes a fine-grained design for the multi-GPU based ptychography.
It parallelizes the ePIE by simultaneously processing all diffraction patterns and updating the 
reconstructed image only once in each iteration. However, this parallelization does not comply with 
the design principle of ePIE and can introduce artifacts while reconstructing experimental data.
Besides, it also suffers from the sub-optimal performance of MPI for intra-node GPU-GPU communications.

The difficulty of scaling applications on multi-GPU configurations is often underestimated.
Simply distributing the workloads and duplicating the single-GPU procedure onto other GPUs usually lead to incorrect 
results and poor scaling. Therefore, implementations on multi-GPU configurations require design considerations 
for communications via mixed topology containing both PCIe and NVlink, asynchronous execution-communication, and so on.
There are limited number of studies that focus on designing fine-grained multi-GPU based implementations. 
BLASX~\cite{wang2016blasx} is a library providing high-performance Level-3 BLAS primitives for the multi-GPU based linear algebra. 
Gunrock~\cite{pan2017multi} and Groute~\cite{ben2017groute} are two frameworks that support the multi-GPU graph analytics.
MAPS-Multi~\cite{ben2015memory} and Blink~\cite{wang2019blink} propose the designs for the multi-GPU based machine learning. 
All these works optimize multi-GPU performance by carefully balancing application workload and tuning GPU P2P communication 
while considering both algorithmic and multi-GPU architectural characteristics. 
They verify that the fine-grained parallelization is essential to fully unleash the multi-GPU's computing power 
for accelerating real-world applications.

\section*{Results and Discussions}
We evaluate both the efficacy and efficiency of our CG-solver based PtyGer (\textbf{Pty}chographic \textbf{G}PU(multipl\textbf{e})-based
\textbf{r}econstruction) tool on a multi-GPU platform. This platform equips two Intel Xeon Silver 4116 CPUs (host) and eight NVIDIA GeForce 
RTX 2080Ti GPUs (devices). 
The host has 768GB total memory and each CPU consists of 12 physical cores (or 24 cores in total for a host). 
Each 2080Ti card is built upon the latest Turing micro-architecture and integrates 68 streaming multiprocessors (SM) with 64 CUDA cores 
per SM, hence totally holds 4352 CUDA cores on chip. Each card has a 11GB off-chip global memory that is shared by all SMs. All 68 SMs 
also share a 5.5MB on-chip L2 cache, and each SM has a 64KB private L1 cache (shared memory). The GPUs are interconnected using PCIe 
that are capable of 11GB/s unidirectional bandwidth. The Turing micro-architecture disables P2P communication via PCIe, hence the GPU-GPU
data transfers are accomplished via host memory.

We evaluate our tool using three dataset: \textit{siemens-star}, \textit{coins}, and \textit{pillar}. The first two 
dataset are synthetic produced by generating the diffraction patterns from a test image (\textit{siemens-star}) 
and a real image (\textit{coins}), respectively. 
The third dataset is a real experimental data from an integrated circuit (IC) sample, and collected at 2-ID-D microscopy beamline at Advanced Photon Source with \SI{.2}{\micro\metre}$\times$\SI{.2}{\micro\metre} beam size (https://www.aps.anl.gov/Microscopy/Beamlines/2-ID-D).
The dimensions of the reconstructed images are $2048\times2048$, $1024\times1024$, and  $1024\times768$, respectively. 
The probe dimensions are $256\times256$ for all dataset.  
We introduce different computation demands for reconstructions by sub-sampling each of the dataset to 4K, 8K, and 16K diffraction patterns.
The total number of iterations is set to 128 for all reconstructions.

\subsection*{Reconstruction efficacy}
\label{sec:efficacy}
In this section, we evaluate the efficacy of PtyGer by examining the accuracy of the reconstructions. 
For each dataset, we calculate the structural similarity (SSIM)~\cite{wang2004image} and 
peak signal-to-noise ratio (PSNR) between the reconstructed image 
and the reference image (ground-truth) after each iteration. 1 in SSIM indicates perfect structural similarity 
while a value larger than 50dB in PSNR typically represents that the difference between two images is very tiny. 
Since the real experimental data 
have no reference object image, we only evaluate the reconstruction results of synthetic datasets. 

\begin{figure}[htb]
\centering
\hspace*{-0.1in}%
\subfigure[][small-scale dataset]{%
\label{fig:siemens-ssim-small}%
\includegraphics[width=0.46\linewidth]{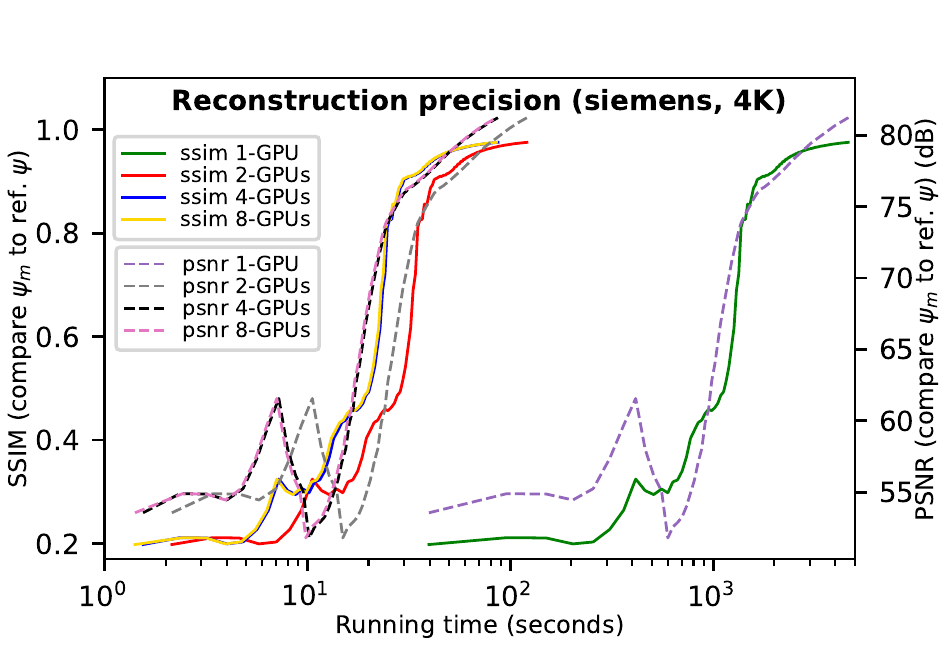}}%
\hspace*{0.02in}%
\subfigure[][medium-scale dataset]{%
\label{fig:siemens-ssim-medium}%
\includegraphics[width=0.46\linewidth]{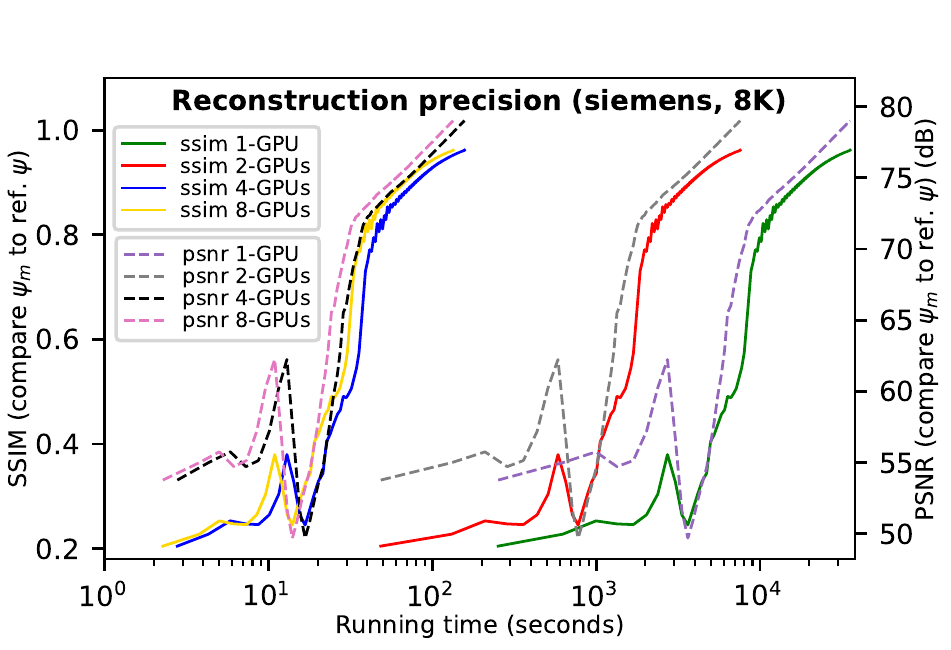}} \\
\vspace*{-0.16in}%
\hspace*{-0.1in}%
\subfigure[][large-scale dataset]{%
\label{fig:siemens-ssim-large}%
\includegraphics[width=0.46\linewidth]{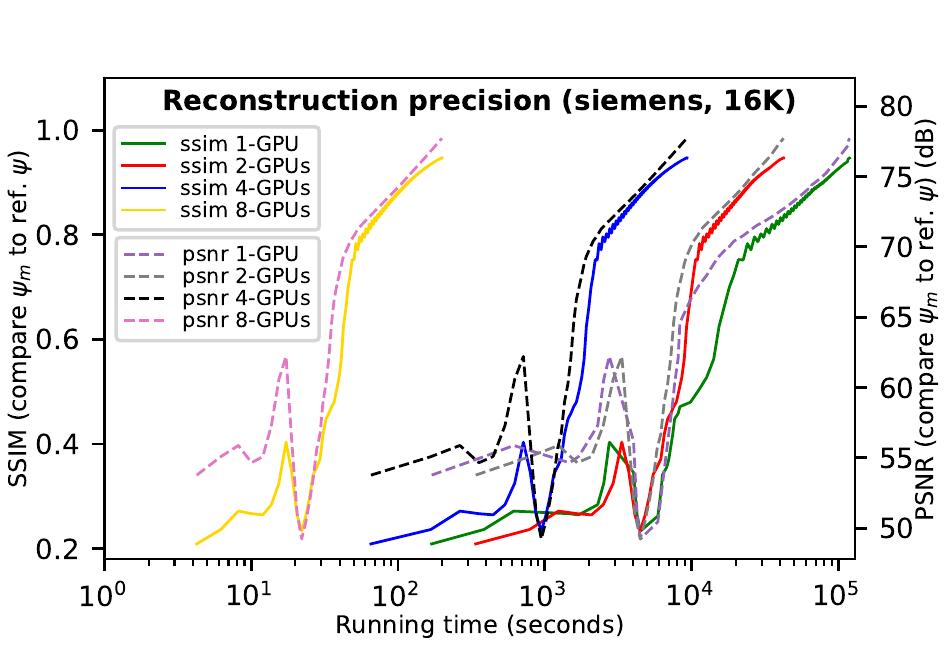}}%
\hspace*{0.08in}%
\subfigure[][medium-scale dataset]{%
\label{fig:siemens-fig}%
\begin{tikzpicture}[spy using outlines={lens={scale=4}, size=1.2cm, connect spies}]
    \node[anchor=south east,inner sep=0] (image) at (0,0){\includegraphics[width=0.46\linewidth]{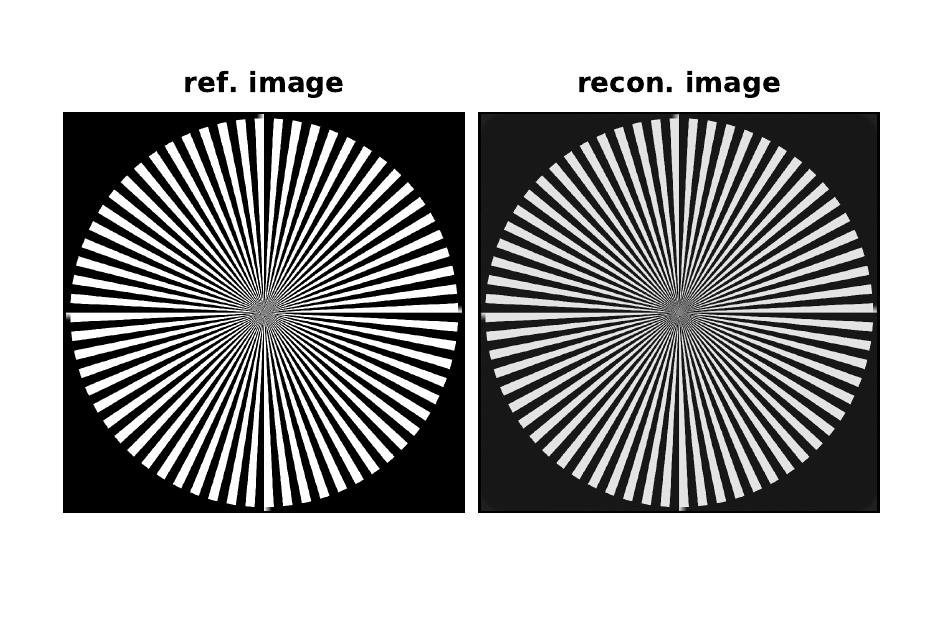}};
    \spy [red] on (-5.78,2.76) in node at (-6.91,1.61);
    \spy [red] on (-2.15,2.76) in node at (-3.28,1.61);
\end{tikzpicture}}%
\caption[]{The reconstruction accuracy evaluations using \subref{fig:siemens-ssim-small} small-scale (4K diffraction patterns), \subref{fig:siemens-ssim-medium} medium-scale (8K), and \subref{fig:siemens-ssim-large} large-scale (16K) synthetic \textit{siemens-star} datasets. For each evaluation, we calculate both the SSIM and PSNR between the reconstructed image and the reference image after each iteration. We use the running time as the x-axis to distinguish the results of different GPU configurations. It is noteworthy that, in each subfigure, the reconstructions with different number of GPUs actually have the same convergence curve. The visual difference is caused by different speeds of running a iteration. \subref{fig:siemens-fig} shows the comparison between the 4-GPU reconstructed image of the 8K dataset and the reference image.}
\label{fig:siemens-ssim}%
\end{figure}

Fig.~\ref{fig:siemens-ssim} demonstrates the reconstruction accuracy of our PtyGer on \textit{siemens-star} datasets using 8-GPU, 
4-GPU, and 2-GPU configurations, respectively. It also includes the single-GPU reconstruction result~\cite{nikitin2019photon} as the counterpart. In each subfigure,
the x-axis indicates the running time of the PtyGer in log-10 scale, while the left-hand y-axis indicates the SSIM and the right-hand y-axis represents PSNR, both between the reconstructed 
image and the reference image. Fig.~\ref{fig:siemens-ssim-small} shows the results on small-scale \textit{siemens-star} dataset 
(4K diffraction patterns). It indicates that although taking different amount of time, the SSIM of all multi-GPU configurations 
approach to reference image while the PSNRs start at around 55dB and reach above 80dB at the end. We also observe that the multi-GPU reconstructions have very similar convergence curves compared to the single-GPU based reconstruction. More specifically, after approximate 
65 iterations, the multi-GPU based reconstructions have their SSIM and PSNR going beyond 0.95 and 75dB and monotonically increasing towards reference image ever since.
Fig.~\ref{fig:siemens-ssim-medium} and Fig.~\ref{fig:siemens-ssim-large} show the results of medium-scale \textit{siemens-star} dataset (8K) and 
large-scale \textit{siemens-star} dataset (16K), respectively. Similarly, they indicate that multi-GPU based reconstructions on both 
datasets have their SSIM converging to 1 and PSNR reaching 80dB and expose the same convergence curves compared to their single-GPU based counterparts.
Fig.~\ref{fig:siemens-fig} compares the actual reconstructed image of the 8K dataset using 4-GPU to the original siemens-star image.
The zoom-in to the center of the images highlight that the hybrid parallelization model and the sub-image alignment in PtyGer introduce no artifacts.
Although reconstructions with different GPU configurations have the same precision, they cost different amount of time to converge. 
The convergence speed is mainly determined by the per GPU workload that can fit into the GPU memory. 
For example, while 4K diffraction patterns fit in 2,4, and 8-GPU configurations, 1-GPU configuration cannot accommodate all the data structures and therefore shows significant slowdown compared to other configurations. 
Similarly, for 16K diffraction patterns, dataset can only fit into 8-GPU configuration, which shows an execution time consistent with 4K diffraction patterns' dataset. However, all the other configurations show performance degradation due to memory limitations.
We will comprehensively evaluate the converging efficiency of PtyGer in next section.

\begin{figure}[htb]
\centering
\hspace*{-0.1in}%
\subfigure[][small-scale dataset]{%
\label{fig:coins-ssim-small}%
\includegraphics[width=0.46\linewidth]{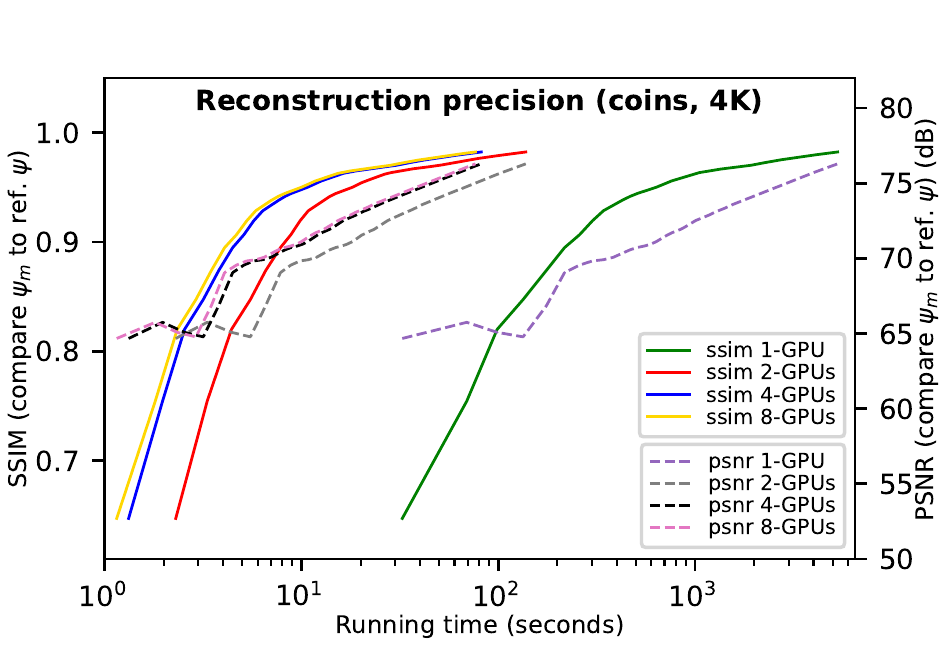}}%
\hspace*{0.02in}%
\subfigure[][medium-scale dataset]{%
\label{fig:coins-ssim-medium}%
\includegraphics[width=0.46\linewidth]{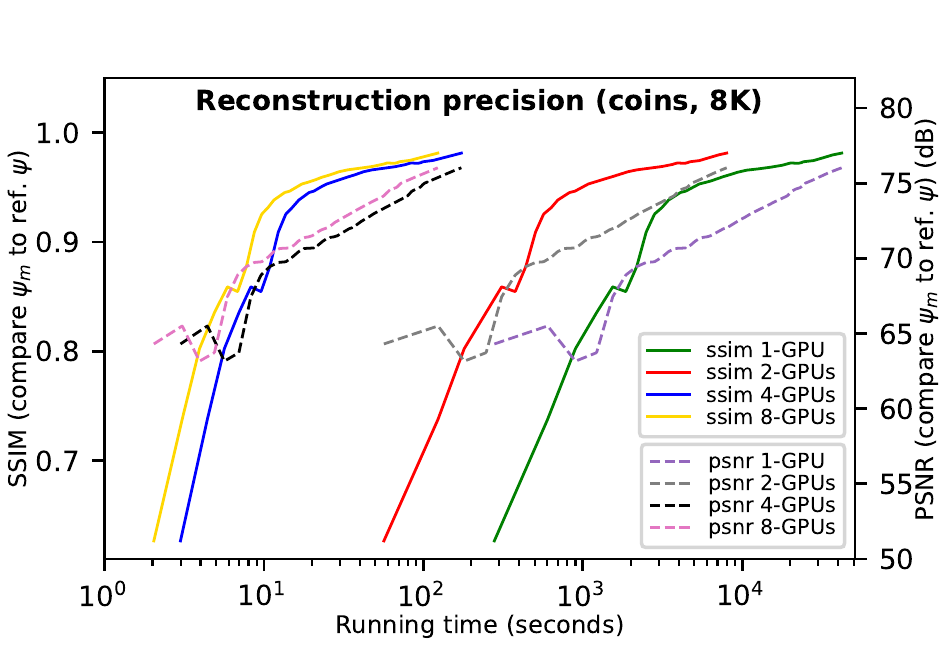}} \\
\vspace*{-0.16in}%
\hspace*{-0.1in}%
\subfigure[][large-scale dataset]{%
\label{fig:coins-ssim-large}%
\includegraphics[width=0.46\linewidth]{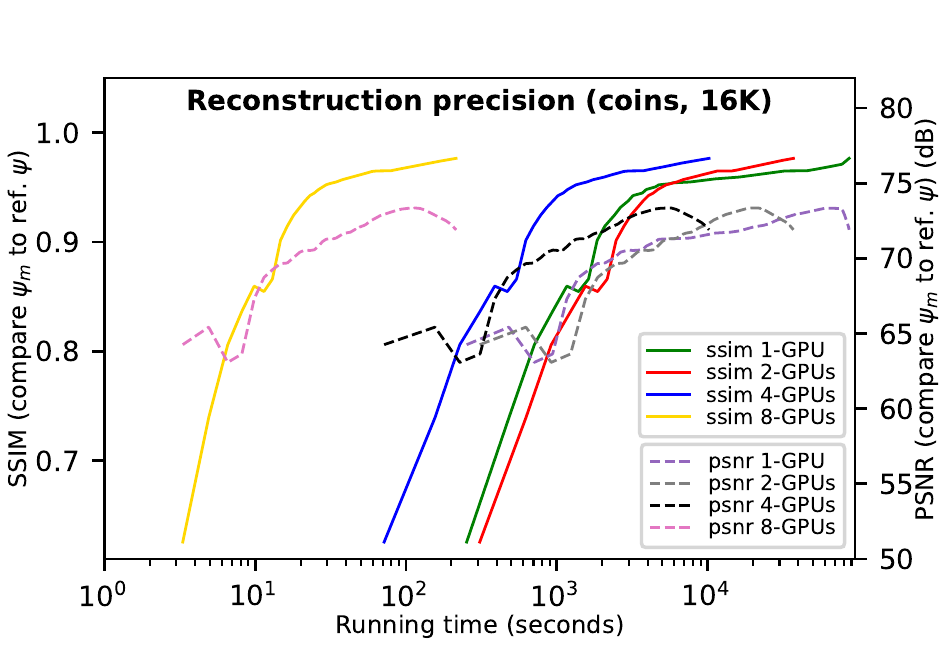}}%
\hspace*{0.08in}%
\subfigure[][medium-scale dataset]{%
\label{fig:coins-fig}%
\begin{tikzpicture}[spy using outlines={lens={scale=4}, size=1.2cm, connect spies}]
    \node[anchor=south east,inner sep=0] (image) at (0,0){\includegraphics[width=0.46\linewidth]{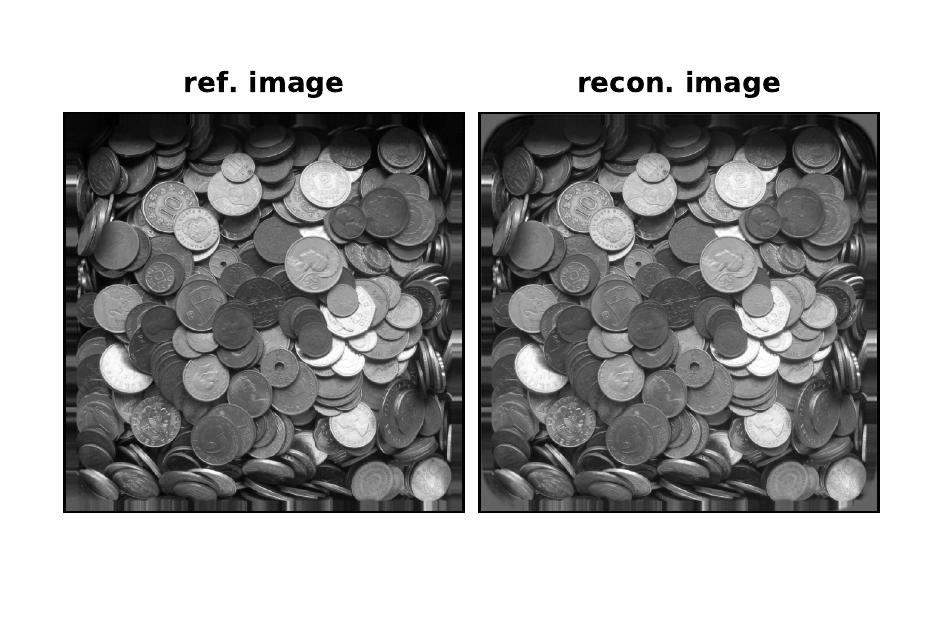}};
    \spy [red] on (-5.78,2.62) in node at (-6.91,1.61);
    \spy [red] on (-2.15,2.62) in node at (-3.29,1.61);
\end{tikzpicture}}%
\caption[]{The reconstruction accuracy evaluations using \subref{fig:coins-ssim-small} small-scale (4K diffraction patterns), \subref{fig:coins-ssim-medium} medium-scale (8K), and \subref{fig:coins-ssim-large} large-scale (16K) synthetic \textit{coins} datasets. For each evaluation, we calculate both the SSIM and PSNR between the reconstructed image and the reference image after each iteration. We use the running time as the x-axis in order to distinguish the results of different GPU configurations. It is noteworthy that, in each subfigure, the reconstructions with different number of GPUs actually have the same convergence curve. The visual difference is caused by different speeds of running a iteration. \subref{fig:coins-fig} shows the comparison between the 4-GPU reconstructed image of the 8K dataset and the reference image.}%
\label{fig:coins-ssim}%
\end{figure}

Fig.~\ref{fig:coins-ssim} demonstrates another set of reconstruction accuracy evaluations using the same 
experimental design as Fig.~\ref{fig:siemens-ssim} on small-scale \textit{coins} (Fig.~\ref{fig:coins-ssim-small}), 
medium-scale \textit{coins} (Fig.~\ref{fig:coins-ssim-medium}), and large-scale \textit{coins} (Fig.~\ref{fig:coins-ssim-large}) 
dataset, respectively. The SSIM and PSNR in 
Fig.~\ref{fig:coins-ssim} indicate that the reconstructions of \textit{coins} datasets with different multi-GPU configurations 
all converge to reference image and the convergence curves are identical to their corresponding 
single-GPU based counterparts'. Fig.~\ref{fig:siemens-fig} displays the actual 4-GPU reconstructed image of the 8K dataset and compares it to the original coins image. We note that no artifacts are observed at the zoomed in centers of the images, where the GPUs exchange border information. 

Above evaluation results show that our PtyGer tool with various number of GPUs can fully preserve the functional 
equivalence compared to the original algorithm implemented on a single GPU. The outputs of our multi-GPU based reconstructions 
are mathematically similar enough to the ground-truths (the reference images), and the convergence trends are identical 
to the original single-GPU based implementation.

\subsection*{Reconstruction efficiency}\label{sec:time}

In this section, we evaluate our PtyGer's reconstruction efficiency, i.e., the speedups compared to the single-GPU 
counterpart~\cite{nikitin2019photon} (the baseline). Last subsection verifies that our PtyGer preserves the algorithmic 
equivalence, hence comparing the convergence speeds of the PtyGer to the counterpart can indicate the performance speedups. 
We measure the convergence speed by recording the least-square norm (2-norm) of the difference between current and 
previous iterations' outputs (i.e., $\psi_{m}-\psi_{m-1}$). 2-norm being 0 represents perfect convergence. 
Practically, when the 2-norm goes below a certain threshold, the reconstruction result is considered converged. 
Notice that our experiments use fixed number of iterations (128) instead of stop criteria, hence the reconstructions keep iterating even if the outputs are converged. 
Therefore, the 2-norms can decrease further or fluctuate within a small range at the end. 
We evaluate the PtyGer's efficiency on all nine datasets.

\begin{figure}[htb]
\centering
\hspace*{0.1in}%
\subfigure[][small-scale dataset]{%
\label{fig:siemens-small}%
\includegraphics[width=0.46\linewidth]{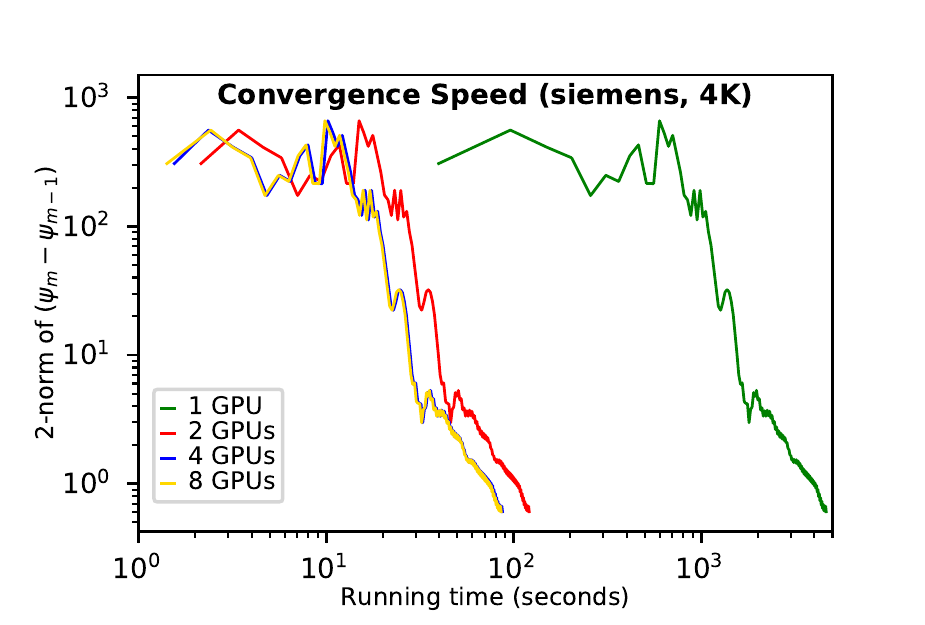}}%
\hspace*{0.02in}%
\subfigure[][time breakdown for small-scale dataset]{%
\label{fig:prof}%
\includegraphics[width=0.46\linewidth]{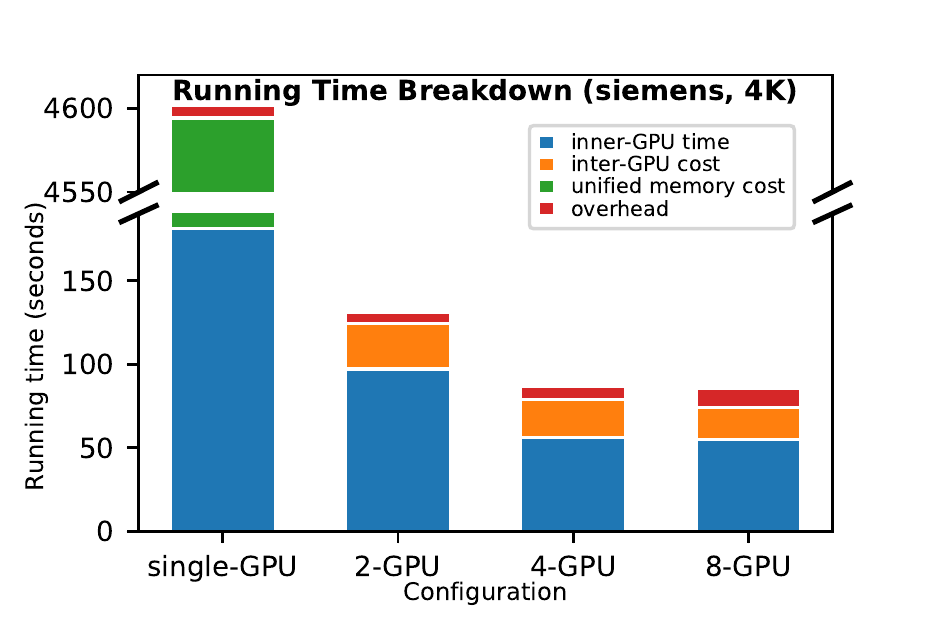}} \\
\vspace*{-0.16in}%
\hspace*{0.1in}%
\subfigure[][medium-scale dataset]{%
\label{fig:siemens-medium}%
\includegraphics[width=0.46\linewidth]{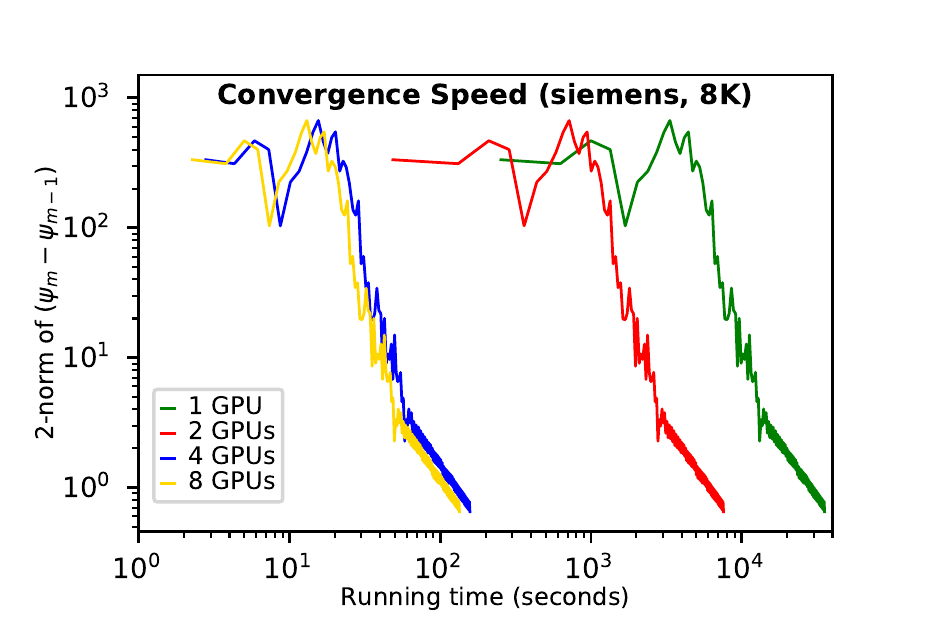}}%
\hspace*{0.02in}%
\subfigure[][large-scale dataset]{%
\label{fig:siemens-large}%
\includegraphics[width=0.46\linewidth]{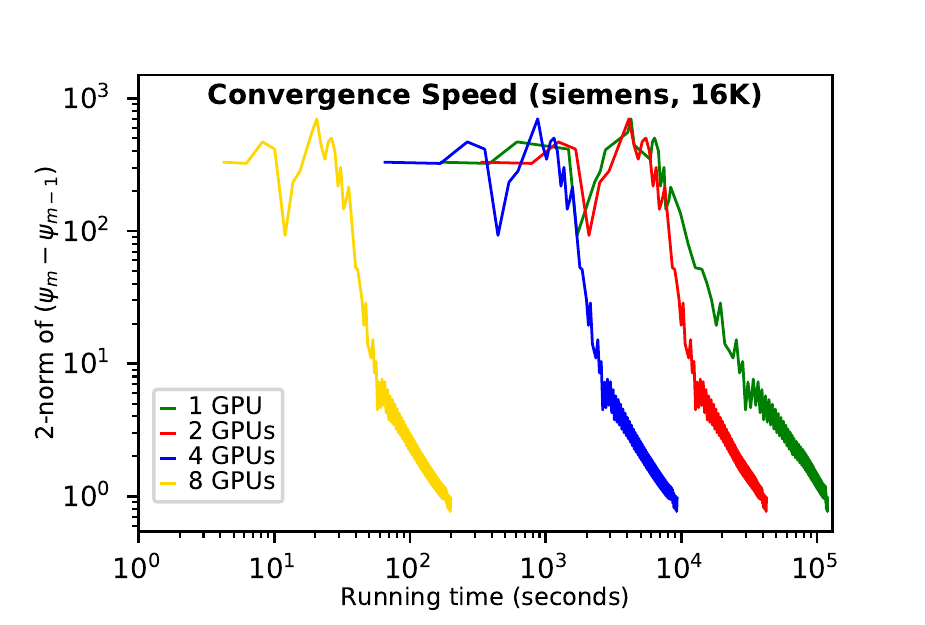}}%
\caption[]{The convergence speed evaluations using \subref{fig:siemens-small} small-scale (4K diffraction patterns), \subref{fig:siemens-medium} medium-scale (8K), and \subref{fig:siemens-large} large-scale (16K) synthetic \textit{siemens-star} datasets. For each evaluation, we apply various GPU configurations (single-GPU, 2-GPU, 4-GPU, and 8-GPU), and compute the difference of the reconstructed images before (i.e., $\psi_{m-1}$) and after (i.e., $\psi_{m}$) Update stage in each iteration by measuring pixels' 2-norm. Notice that we use log-10 scale for both axes. \subref{fig:prof} provides more details for \subref{fig:siemens-small} by breaking down the running time. In single-GPU, notice that there is no inter-GPU cost and we use a broken bar for unified memory cost for better illustration.}%
\label{fig:siemens}%
\end{figure}

Fig.~\ref{fig:siemens} demonstrates the reconstruction convergence rates of three \textit{siemens-star} datasets using 8, 4, 2 and 1 GPU configurations. In each subfigure, the x-axis indicates the running time and the y-axis indicates the 2-norm. 
We emphasize that both axes are in log-10 scale. Consequently, the lower bound 
of 2-norm is negative infinity (i.e., 0 in regular scale). Fig.~\ref{fig:siemens-small} shows the efficiency comparisons of 
reconstructions on small-scale \textit{siemens-star} dataset (4K diffraction patterns). 
Fig.~\ref{fig:prof} breaks down the running time under each configuration. We use a broken bar to demonstrate 
the outlier (i.e., single-GPU). As 
indicated in both figure, single-GPU based reconstruction takes approximately 1.3 hours. 
This slow execution is caused 
by that the size of dataset exceeds the memory capacity of a single GPU. We note that the CG-solver yields large memory footprint 
during the execution due to buffering the intermediate results. Instead aborting the execution, the reconstruction 
leverages CUDA unified memory to handle the excess data. It provides the unified virtual address (UVA) space to the GPU and CPU memories, 
so the program can automatically utilize the CPU memory to assist the GPU execution. However, this technique has a large overhead. 
As shown in Fig.~\ref{fig:prof}, unified memory cost (green segment) dominates the single-GPU execution time (95.8\%). 
The cost includes expensive CPU-GPU data migrations and page fault handlings occurring at each CUDA arithmetic operation.  
Fig.~\ref{fig:siemens-small} shows the 2-GPU based reconstruction is 38.6X faster than the original single-GPU reconstruction. 
The speedup is \emph{superlinear} due to that the splitting of dataset makes the workload of each GPU (2K diffraction patterns per GPU) fit 
into GPU memory hence the 2-GPU reconstruction avoids host-device communication. Fig.~\ref{fig:prof} shows that inner GPU 
computing (blue segment) takes the majority (74.2\%) of the 2-GPU reconstruction time. One other main component (20.8\%) is the 
inter-GPU cost (orange segment) including GPU-GPU synchronizations and communications. The one-time overhead (red segment) 
consisting of data preparation and device management is minor (5\%). Fig.~\ref{fig:siemens-small} also shows that the 
4-GPU based reconstruction has 1.5X speedup compared to the 
2-GPU reconstruction. We observe that in Fig.~\ref{fig:prof}, although the inner-GPU computing of the 4-GPU is nearly 2X faster due 
to the halved per GPU workload, the overall speedup is sublinear. 
The main reason of this is since more number of devices is involved in computation, the inter-GPU cost and one-time overhead are slightly larger than the 2-GPU. 
Fig.~\ref{fig:siemens-small} shows the 8-GPU based reconstruction has nearly the same efficiency as the 4-GPU configuration. 
This is because increasing the number of GPUs makes each GPU's workload (0.5K diffraction patterns) is too small to efficiently occupy 
the GPU computing resources. 
Fig.~\ref{fig:prof} shows that both 4-GPU and 8-GPU configurations are memory bound since they have similar computation times.

Fig.~\ref{fig:siemens-medium} and \ref{fig:siemens-large} show the reconstruction efficiency comparisons on 
medium-scale \textit{siemens-star} dataset and large-scale \textit{siemens-star} dataset, respectively. In Fig.~\ref{fig:siemens-medium}, 
we observe that, even though running faster than single-GPU configuration, the 2-GPU reconstruction nevertheless 
takes very long time since the divided workload on each GPU (4K per GPU) still exceeds the GPU memory capacity. 
We also observe that the 4-GPU configuration is 42.4X faster than the 2-GPU. 
The speedup is superlinear due to the same reason as mentioned before. 
The 8-GPU based reconstruction is 1.3X faster than the 4-GPU. 
Compared to the 8-GPU based reconstruction on the small-scale \textit{siemens-star} dataset, the reconstruction on the medium-scale dataset doubles the workload per GPU and therefore better utilizes the computing resources. 
Thus, for the medium-scale dataset, increasing the GPU number from four to eight can improve the overall performance. Fig.~\ref{fig:siemens-large} shows 
that, for the large-scale \textit{siemens-star} dataset, even the 4-GPU configuration has insufficient memory. 
On the other hand, the 8-GPU configuration makes the per GPU workload small enough to fit into the GPU memory and achieves superlinear speedup (41.5X) against the 4-GPU configuration. 

\begin{figure}[htb]
\centering
\hspace*{0.1in}%
\subfigure[][small-scale dataset]{%
\label{fig:coins-small}%
\includegraphics[width=0.46\linewidth]{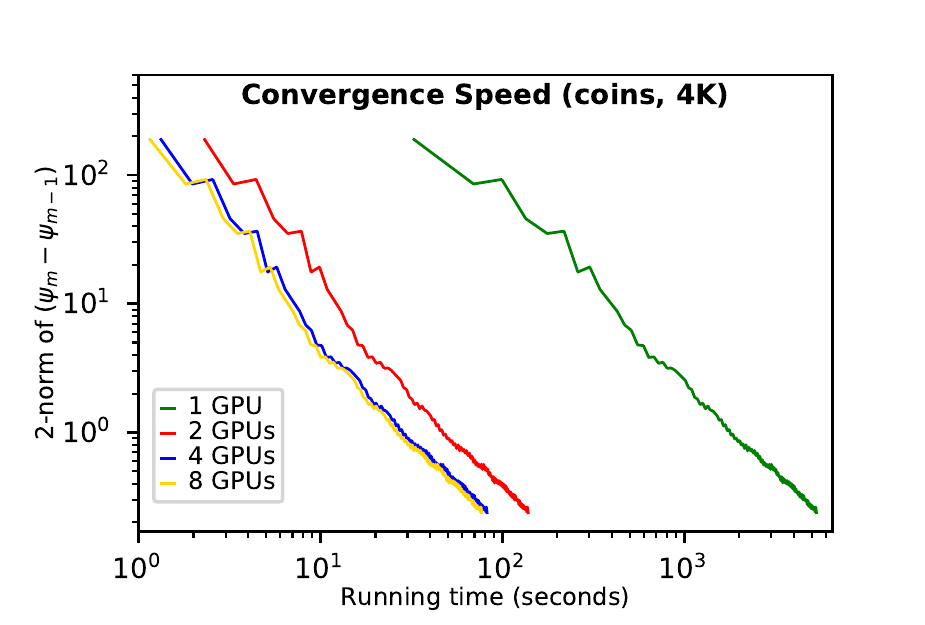}}%
\hspace*{0.02in}%
\subfigure[][medium-scale dataset]{%
\label{fig:coins-medium}%
\includegraphics[width=0.46\linewidth]{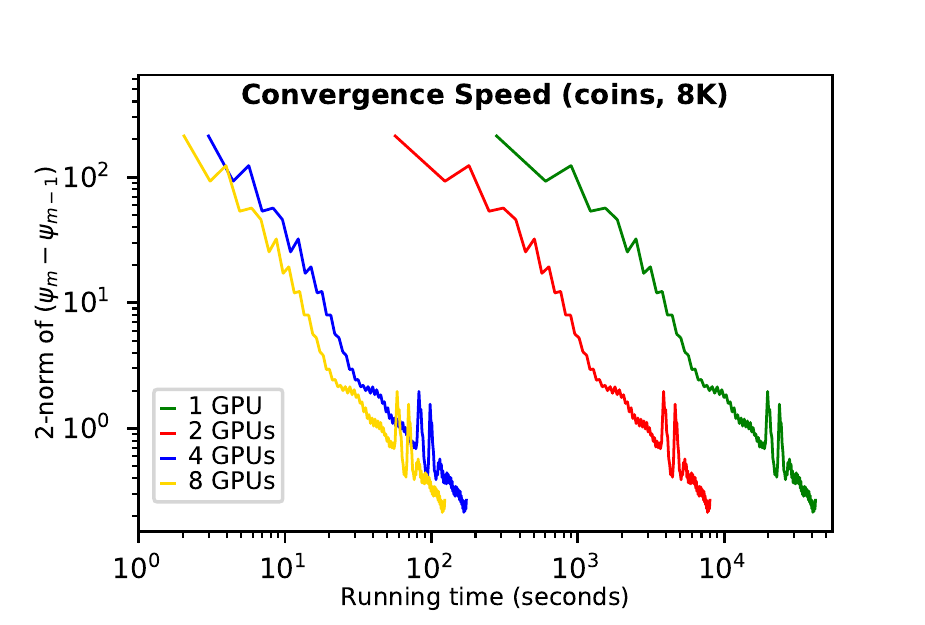}} \\
\vspace*{-0.16in}%
\subfigure[][large-scale dataset]{%
\label{fig:coins-large}%
\includegraphics[width=0.46\linewidth]{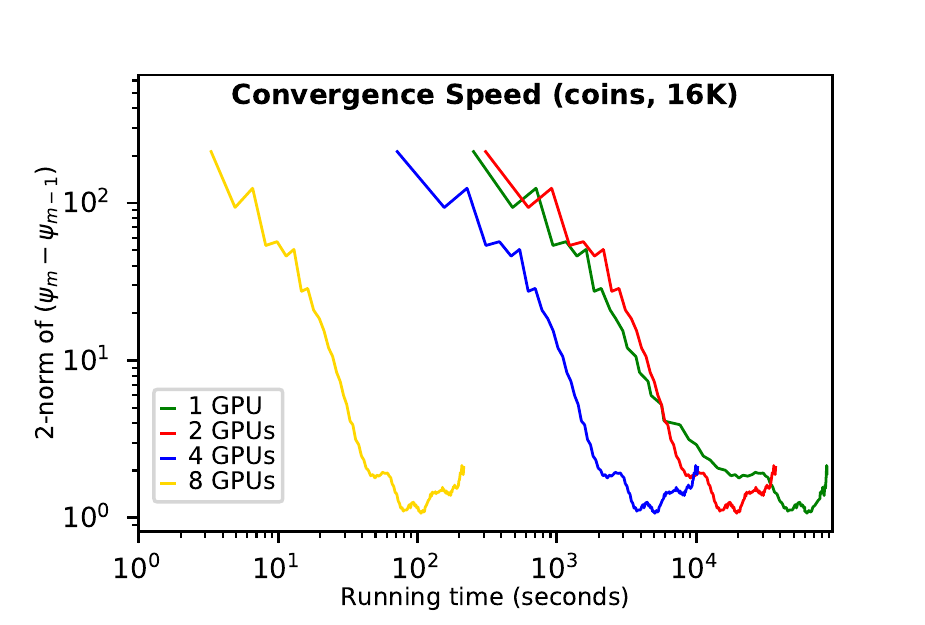}}%
\caption[]{The convergence speed evaluations using \subref{fig:coins-small} small-scale (4K diffraction patterns), \subref{fig:coins-medium} medium-scale (8K), and \subref{fig:coins-large} large-scale (16K) synthetic \textit{coins} datasets. For each evaluation, we apply various GPU configurations (single-GPU, 2-GPU, 4-GPU, and 8-GPU), and compute the difference of the reconstructed images before (i.e., $\psi_{m-1}$) and after (i.e., $\psi_{m}$) Update stage in each iteration by measuring pixels' 2-norm. Notice that we use log-10 scale for both axes.}%
\label{fig:coins}%
\end{figure}

\begin{figure}[htb]
\centering
\hspace*{-0.1in}%
\subfigure[][small-scale dataset]{%
\label{fig:pillar-small}%
\includegraphics[width=0.46\linewidth]{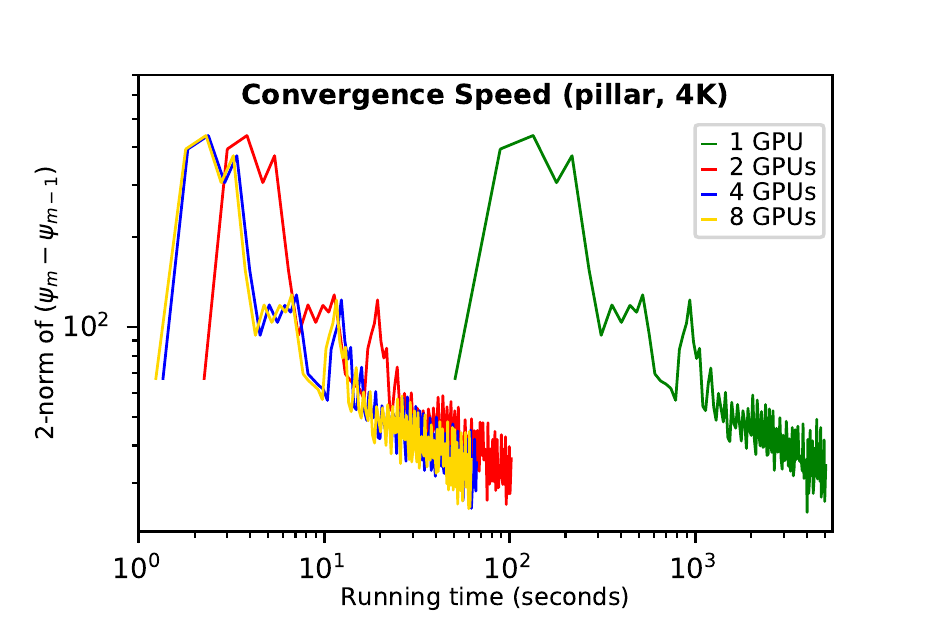}}%
\hspace*{0.02in}%
\subfigure[][medium-scale dataset]{%
\label{fig:pillar-medium}%
\includegraphics[width=0.46\linewidth]{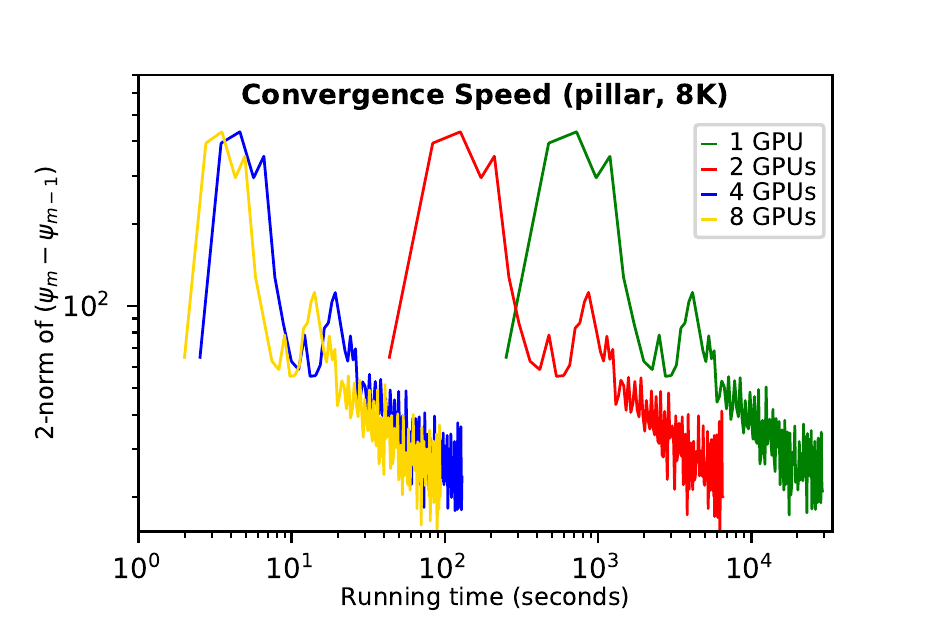}} \\
\vspace*{-0.16in}%
\hspace*{-0.1in}%
\subfigure[][large-scale dataset]{%
\label{fig:pillar-large}%
\includegraphics[width=0.46\linewidth]{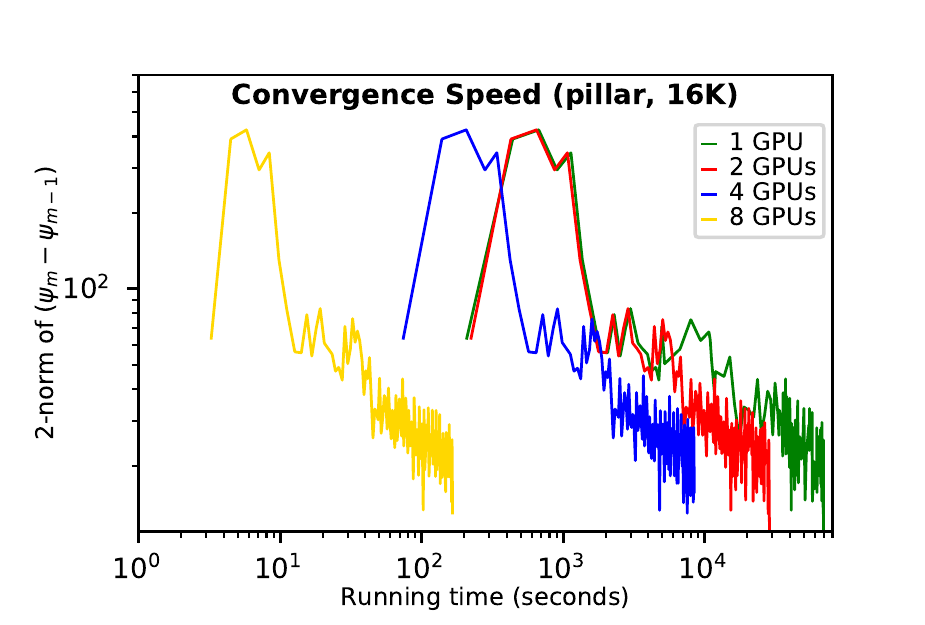}}%
\hspace*{-0.05in}%
\subfigure[][medium-scale dataset]{%
\label{fig:pillar-fig}%
\begin{tikzpicture}[spy using outlines={lens={scale=3}, size=1.2cm, connect spies}]
    \node[anchor=south east,inner sep=0] (image) at (0,0){\includegraphics[width=0.47\linewidth]{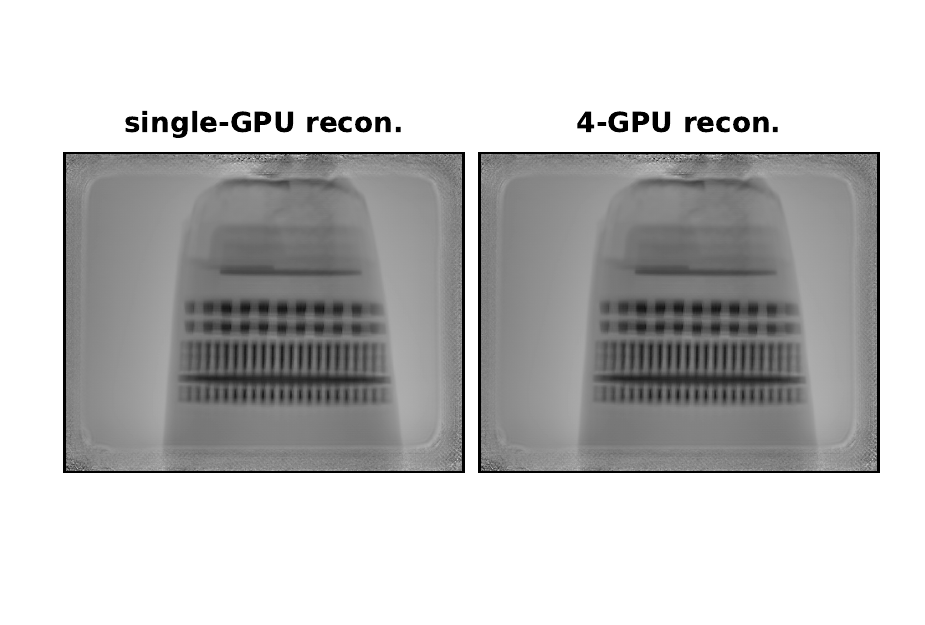}};
    \spy [red] on (-5.86,2.66) in node at (-7.09,1.98);
    \spy [red] on (-2.15,2.66) in node at (-3.37,1.98);
\end{tikzpicture}}%
\caption[]{The convergence speed evaluations using \subref{fig:pillar-small} small-scale (4K diffraction patterns), \subref{fig:pillar-medium} medium-scale (8K), and \subref{fig:pillar-large} large-scale (16K) real-beamline \textit{pillar} datasets. For each evaluation, we apply various GPU configurations (single-GPU, 2-GPU, 4-GPU, and 8-GPU), and compute the difference of the reconstructed images before (i.e., $\psi_{m-1}$) and after (i.e., $\psi_{m}$) Update stage in each iteration by measuring pixels' 2-norm. Notice that we use log-10 scale for both axes. \subref{fig:pillar-fig} compares the 4-GPU reconstructed image of 8K dataset to the single-GPU result.}%
\label{fig:pillar}%
\end{figure}

Fig.~\ref{fig:coins} and Fig.~\ref{fig:pillar} show another two sets of efficiency comparisons on \textit{coins} 
and \textit{pillar} datasets, respectively. 
As shown in Fig.~\ref{fig:coins-small} and Fig.~\ref{fig:pillar-small}, for the small-scale datasets, the workloads exceed the capacity of a single GPU memory hence the single-GPU based reconstruction is very slow. 
The 2-GPU configurations accommodate the per GPU workload within the GPU memory capacity thus achieve superlinear speedups (38.4X in Fig.~\ref{fig:coins-small} and 39.7X Fig.~\ref{fig:pillar-small} compared to the single-GPU one). 
Increasing the number of GPUs from two to four can achieve additional 1.7X and 1.6X speedups. 
However, further increasing the number of GPUs from four to eight shows no improvement due to the underutilized GPU resources. 
For the medium-scale datasets, the 2-GPU configuration cannot keep the per GPU workload within the GPU memory capacity hence 4-GPU configuration shows 40.2X (Fig.~\ref{fig:coins-medium}) and 42.4X (Fig.~\ref{fig:pillar-medium}) superlinear speedups. 
The 8-GPU configuration can further improve the efficiency by 1.4-fold (Fig.~\ref{fig:coins-medium}) 
and 1.4-fold (Fig.~\ref{fig:pillar-medium}) compared to the 4-GPU. For the large-scale datasets, the per GPU workload 
exceeds the GPU memory capacity even with four GPUs, but the 8-GPU based reconstructions can still afford the large 
workloads and achieve superlinear speedups (42.9X (Fig.~\ref{fig:coins-large}) and 43.1X (Fig.~\ref{fig:pillar-large}), respectively) 
over the 4-GPU configurations. Fig.~\ref{fig:pillar-fig} displays the reconstructed pillar image with 8K diffraction patterns using 4-GPU. Since there is no reference pillar image, we compare the 4-GPU reconstructed image to the single-GPU result. 
The comparison shows that scaling the reconstruction of real-beamline data to multiple GPUs using PtyGer preserves the reconstructed image's quality.
Focusing on the image centers shows that there are no artifacts at the overlapped regions among GPUs.
All above comparisons verify that our multi-GPU PtyGer can successfully increase the overall reconstruction efficiency. 
The rate of increase is affected by both dataset size and the number of GPU. 
If the workload exceeds the GPU memory capacity, then adding more number of GPUs can achieve superlinear speedup as aggregated GPUs memory decreases the host-device memory operations.
Once the workload fits into aggregated memory, adding more GPUs can still improve the reconstruction efficiency, however the performance improvement is sublinear due to inter-GPU synchronization and communication overheads.
However, once the per GPU workload is significantly insufficient to occupy the GPU computing resources, the performance gain stops growing. 

\subsection*{Summary and discussion}
We comprehensively evaluate our PtyGer using both synthetic and real-beamline data with various sizes. 
The evaluation of the efficacy shows that PtyGer fully preserves the accuracy of the original sequential algorithm. 
We also show great scalability efficiency with PtyGer, in some cases superlinear, with increasing number of GPUs and large workloads.
The multi-GPU based reconstructions are significantly faster than the  single-GPU counterparts due to the elimination of unified memory usage (host-device data movement) and aggregated GPU memory. 
For the multi-GPU, when the per GPU workload fits into GPU memory, doubling the number of device results in up to 1.7X performance improvement. 
The sublinearity is caused by the unavoidable preprocessing overhead and inter-GPU costs. Our hybrid parallelization model design has minimized the inter-GPU communication.

The optimal number of GPU devices depends on the dataset size. More GPUs do not necessarily provide better performance.
When the per GPU workload is too small, system reaches the memory I/O bound. 
Therefore, adding more GPUs will not increase the reconstruction speed and lead to a waste of the computational resources. 
Empirically, using 2 GPUs for a small dataset can achieve the best resource utilization; 4 GPUs are adequate for a medium dataset; 8 GPUs are needed for a large dataset. 
Whenever sufficient number of GPUs is utilized for a given workload, any excess GPUs can be used for reconstructing other view angles in 3D ptychography.

%
%
%
%
%
%
%

\section*{Methods}

\subsection*{Background}\label{sec:backgrounds}
In this section, we provide the background information about the ptychography model, the iterative phase-retrieval algorithm, and the gradient-based solvers.

\subsubsection*{Forward problem of ptychography}\label{subsec:model}
\begin{figure}[htb]
\centering
\includegraphics[width=0.65\linewidth]{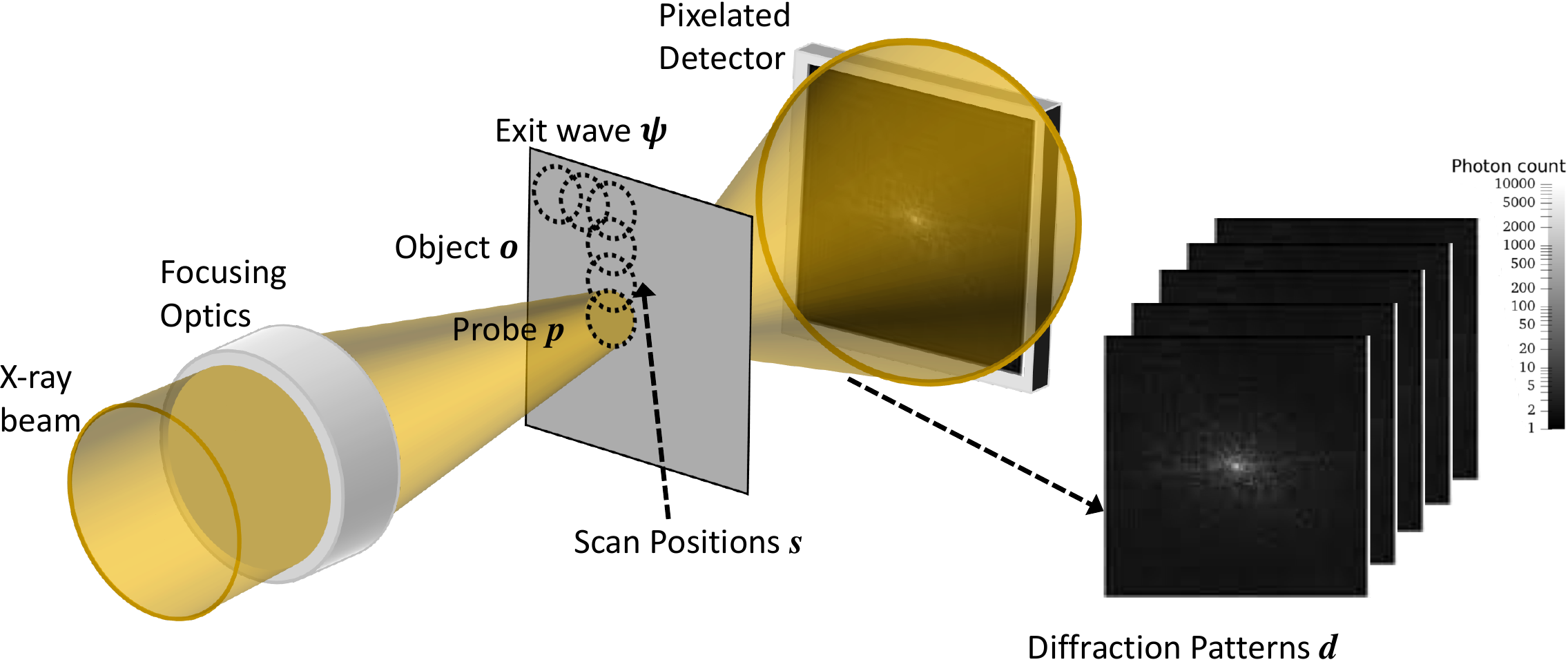}
\caption{A typical setup of ptychography data collection. The focused probe $p$ raster-scans
the object $o$ at overlapped positions $s$. A series of diffraction patterns of the exit waves
are collected by the pixelated detector to form the diffraction data $d$.}
\label{fig:ptycho_model}
\end{figure}

Fig.~\ref{fig:ptycho_model} schematically shows a generic ptychography setup. In ptychography, a focused coherent x-ray beam of light (i.e., 
probe \textit{p}) is used to scan the object \textit{o}, and the diffraction patterns of the transmitted radiation is collected via 
a pixelated photon counting detector. The probe scans the object in a series of consecutive and overlapped illumination positions 
(i.e., scan position \textit{s} in Fig.~\ref{fig:ptycho_model}). Accordingly, the detector collects a series of diffraction patterns denoted 
by \textit{d}. Ptychography can be mathematically described by the forward model from exit wave $\psi$ to diffraction pattern \textit{d}: 
\begin{equation}\label{Eq:fwd_model}
  |\Gop \psi|^2 = |\Fop \Qop\psi|^2 = d,
\end{equation}
where $\Gop$ is the ptychography operator that represents applying Fourier transforms $\Fop$ to the diagonal operator $\Qop$.
$\Qop$ acts as an element-wise multiplication of probe \textit{p} and exit wave $\psi$ at all scan positions.

\subsubsection*{Inverse problem for ptychography}\label{subsec:inverse}

In this section we briefly recapitulate our approach for solving 2D ptychography problem. For details we refer to our former papers \cite{nikitin2019photon,aslan2019joint}, where we considered Maximum-Likelihood (ML) and Least-Squares (LS) estimates for the solution. As an example, here we consider the ML estimate formulation, although all the techniques introduced in this paper are also applicable to the LS estimator.

The exit wave \textit{$\psi$} can be numerically reconstructed based on the solution of the phase retrieval problem.
Photon collection of the detector is a Poisson process~\cite{Reiffen:63}, and the probability of acquiring data $d_j$, $j=1,2,\dots,n$, is given by the likelihood function.

The corresponding maximum likelihood (ML) estimate of $\psi$ then can be computed by solving  the following minimization problem
\begin{equation}\label{Eq:ML}
    \begin{aligned}
        F(\psi)=\sum_{j=1}^{n} \left\{|\Gop \psi|_j^2-2d_j\log |\Gop \psi|_j \right\}\to \min,
    \end{aligned}
\end{equation}
for which the gradient is given as follows
\begin{equation}\label{Eq:gradient}
  \nabla_\psi F(\psi) =\Gop^H\left(   \Gop\psi-\frac{d}{(\Gop\psi)^*}\right),
\end{equation}
where $\Gop^H=\Qop^H\Fop^H$ and $\Qop^H$ operates as an element-wise multiplication by the conjugate of probe \textit{p}, and $\Fop^H$ 
is the inverse Fourier transform.

With the gradient, we can iteratively reconstruct the object image using various solvers. We can use the gradient-descent (GD) solver to reconstruct the image~\cite{aslan2019joint}. In this case, iterations of updating estimated object can be written as follows
\begin{equation}\label{Eq:update2}
  \psi_{m+1} = \psi_{m} - \gamma \nabla_\psi F(\psi)
\end{equation}
where $\gamma$ is given as a constant small step length.
The GD solver has a slower convergence rate compared to conjugate-gradients (CG) family of solvers \cite{DaiYuan:99,Dai:00,Polak:69,Polyak:69}. In our former work the CG solver with the Dai-Yuan conjugate direction \cite{DaiYuan:99} has demonstrated a faster convergence rate in 2D ptychography compared to analogues. The iterations for object updating with the Dai-Yuan direction can be written as:
\begin{equation}\label{Eq:update}
  \psi_{m+1} = \psi_m + \gamma_m \eta_m.
\end{equation}
where $\gamma_m$ is the step length and $\eta_m$ is the search direction given by the recursive Dai-Yuan~\cite{DaiYuan:99} formula:
\begin{equation}\label{Eq:DaiYuan}
\begin{aligned}
  \eta_{m}= -\nabla_\psi F(\psi_{m})+\frac{\norm{\nabla_\psi F(\psi_{m})}_2^2}{ \langle \eta_{m-1},{\nabla_\psi F(\psi_{m})}-{\nabla_\psi F(\psi_{m-1})\rangle }}\eta_{m-1}
  \end{aligned}
\end{equation}
with the initial direction as the steepest descent direction, $\eta_0=-\nabla_\psi F(\psi_0)$. $\langle a,b\rangle$ is defined as $\sum_i^z a^*_i b_i$ and $z$ is the number of pixels of object $\psi$.
The step length $\gamma_m$ is computed through the line-search procedure~\cite{nocedal2006numerical}.
Line-search starts at an initial large step length, and repeatedly shrinks it
until the following is satisfied,
\begin{equation}\label{Eq:linesearch}
  F(\psi_m+\gamma_m\eta_m) \leqslant F(\psi_m) + \gamma_m t,
\end{equation}
where $t$ is a constant termination parameter usually set to 0 in practice.

\begin{figure}[htb]
\centering
\includegraphics[width=0.75\linewidth]{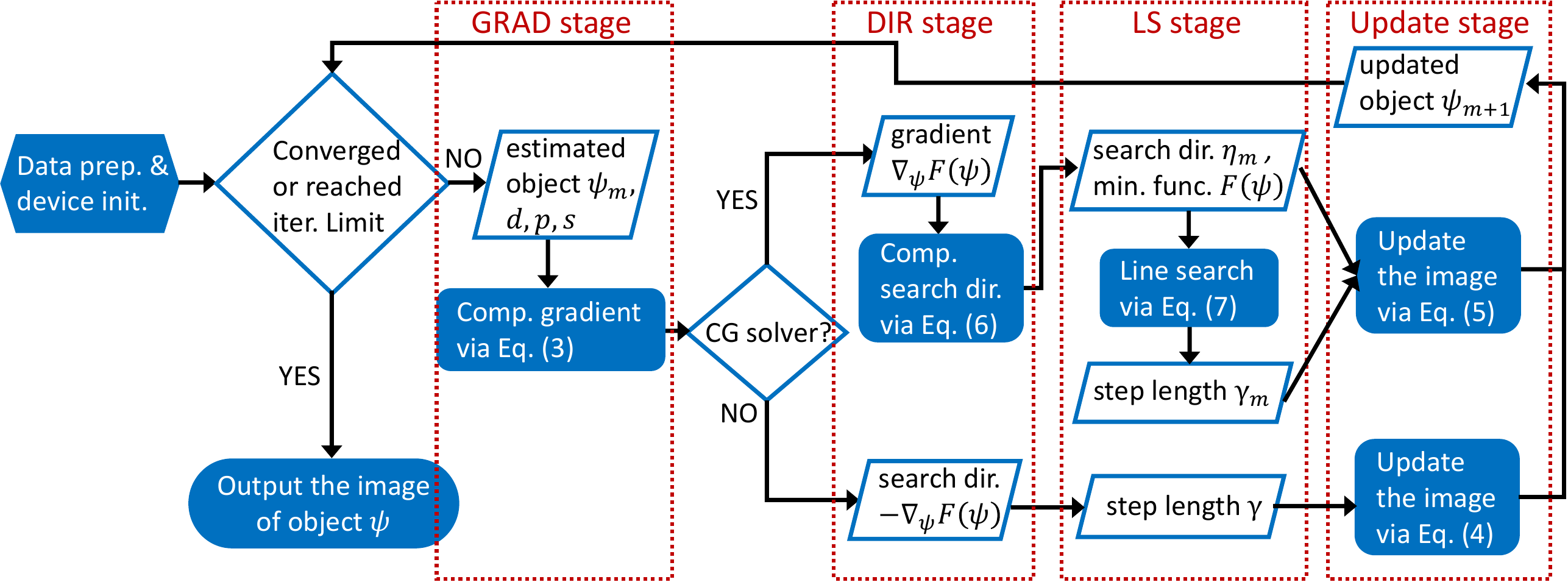}
\caption{The flowchart of iterative maximum-likelihood phase-retrieval algorithm using the gradient-based solver. Each iteration can be 
divided into four stages: \textit{GRAD stage}, \textit{DIR stage}, \textit{LS stage}, and \textit{Update stage}. }
\label{fig:workflow}
\end{figure}

The complete iterative solution of the ptychographic reconstruction can be summarized with the 
workflow shown in Fig.~\ref{fig:workflow}. In the workflow, an iteration is divided to four stages, where 
we apply different parallelization schemes to different stages. The details of our parallelization schemes are elaborated in next section.

\subsection*{Multi-GPU based parallelization designs}\label{sec:design}
In this section, we elaborate the design of our novel parallelization model and PtyGer tool. We start with 
the introduction of a general parallelization design and discuss its limitations for the CG solver.
We then present our hybrid parallelization model that can overcome the two challenges of CG solver parallelization.

\subsubsection*{General parallelization design}\label{subsec:design1}
A general multi-GPU parallelization design consists of \textit{\textbf{workload distribution}} 
and \textit{\textbf{inter-GPU communication}}, which has been employed by some existing works, 
e.g., PtychoLib~\cite{nashed2014parallel}. The objectives of these two components are performing 
diffraction pattern distribution and sub-image border exchanges between GPUs, respectively.

\begin{figure}[htb]
\centering
\includegraphics[width=0.7\linewidth]{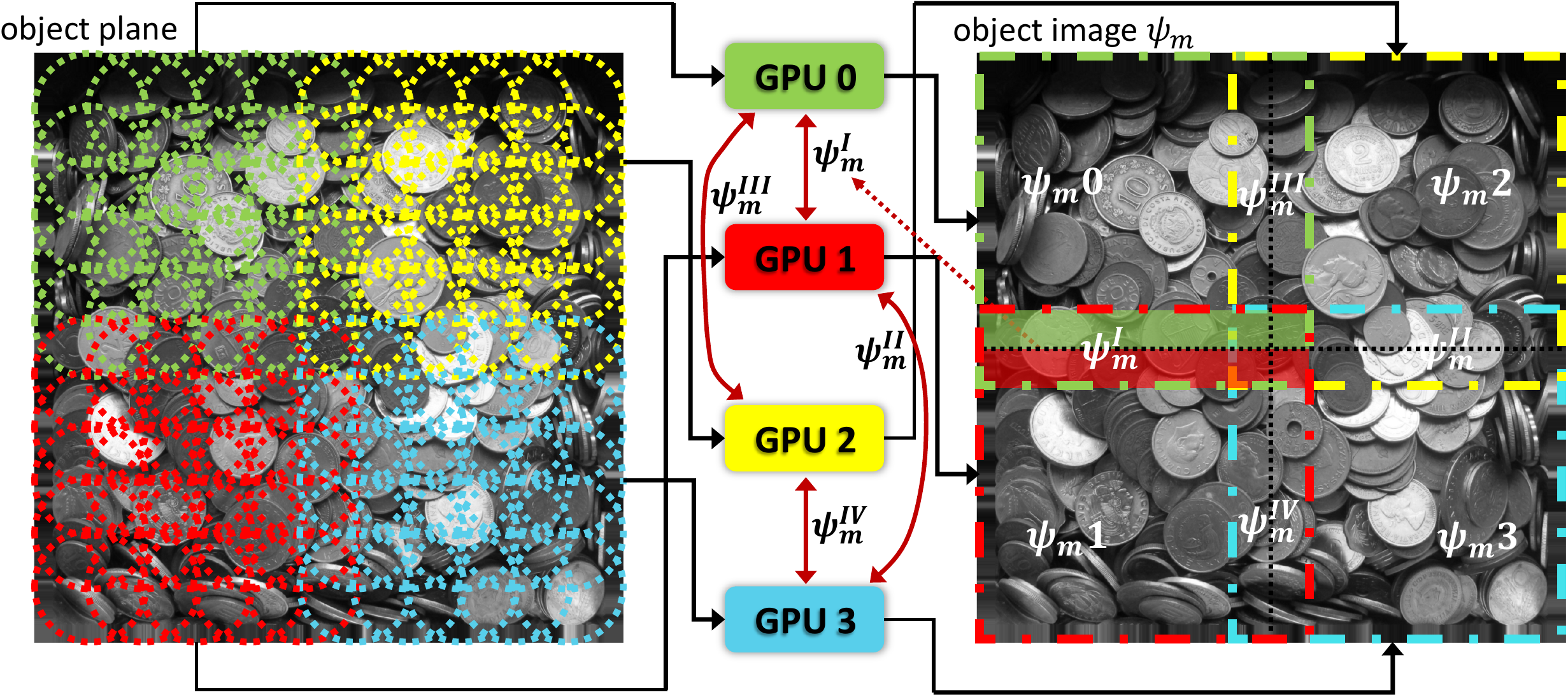}
\caption{The basic parallelization design for a 4-GPU platform. The left-hand side of the GPUs illustrates the workload distribution 
while the right-hand side demonstrates the inter-GPU sub-image border exchange. Each GPU stores a portion of the diffraction patterns 
and locally reconstructs the corresponding sub-image. Neighbor GPUs exchange the sub-image borders after each iteration.}
\label{fig:basic}
\end{figure}

\textit{\textbf{Workload distribution:}} 
In ptychography, the size of the diffraction patterns \textit{d} defines the workload size. 
Suppose that the detector is $512 \times 512$ and the number of scan positions is 16K, then \textit{d} requires $512^2\cdot16K\cdot4=16$GB memory space. 
This memory requirement, together with other data structures needed for reconstruction, can easily exceed available memory in commodity GPUs.
Therefore, we partition the  diffraction patterns \textit{d} and distribute them to many GPUs. 
The left-hand side of Fig.~\ref{fig:basic} schematically demonstrates the workload distribution scheme. 
On the left object plane, each circle represents a scan position and has a corresponding diffraction pattern (omitted in the figure for clarity). 
The diffraction patterns are divided and distributed to the GPUs, then each GPU locally reconstructs the corresponding sub-image as shown on the right object image. 
For example, the diffraction patterns collected at green scan positions are stored in GPU 0, hence GPU 0 can locally reconstruct sub-image $\psi_m0$. 
Notice that to keep the sub-image borders accurate, in each GPU, we store additional diffraction patterns at the halo of the sub-image~\cite{guizar2014high}, and 
exchange the sub-image borders with neighbor GPUs after each iteration. 

\textit{\textbf{Inter-GPU communication:}} 
To handle the border blurring, each GPU reconstructs also the halo of the corresponding sub-image hence has overlaps with its neighbor GPUs.
The right-hand side of Fig.~\ref{fig:basic} shows how the sub-images intersect with their neighbors. 
For each sub-image, local reconstructions only guarantee that the central area (the area within the black dotted lines) is accurate.
It sub-image halos are blurred due to the missed diffraction patterns that stored in the neighbor GPUs.
Therefore, the border areas of the sub-images should be exchanged between GPUs after each object-update iteration to compensate the halo blurring.
For instance, GPU 0 and GPU 1 reconstruct two sub-areas enclosed by the green and red dashed squares, respectively. After each object update, 
the reconstructed sub-image $\psi_m0$ and  $\psi_m1$ are accurate compared to the whole image update. However, the overlapped area $\psi_m1$
is inaccurate in both GPUs. Specifically, the green area is part of $\psi_m1$'s halo and is blurred in GPU 1 while the red area is blurred in GPU 0.
To make the halos accurate, GPU 0 then transfers the accurate green area to GPU 1 and receives the accurate red area transferred from GPU 1. 

The general parallelization of ptychography can be summarized as follows: 
it partitions the diffraction patterns and distributes them to different GPUs in accordance with the GPU capacity.
At each iteration, GPUs locally reconstruct the corresponding sub-images with the halos, then synchronize 
and exchange the borders with their neighbor GPUs. 
After a certain number of iterations or reaching to a stop criterion, we align the locally reconstructed sub-images to obtain the entire object image. 

\subsubsection*{Challenges of CG solver parallelization}\label{subsec:challenges}

Although the general parallelization is fully compatible with some reconstruction algorithms, 
e.g., ePIE and GD solver (Eq.~\eqref{Eq:update2} is directly applicable to partial reconstructions), 
it fails to preserve the CG solver's algorithmic equivalence due to two main challenges.\\

\noindent
\textit{\textbf{Challenge 1:}} In each iteration, the CG solver computes the search direction $\eta_{m}$ using Eq.~\eqref{Eq:DaiYuan}.
Since $\eta$ and $\nabla_\psi F(\psi)$ are matrices that have the same shape as the object image, the partial processing is compatible 
with the first term of Eq.~\eqref{Eq:DaiYuan}. However, the partial processing is not directly applicable to the second term's coefficient computation.

We denote the coefficient as 
\begin{equation}\label{eq:cg_ch1_1}
\begin{aligned}
  \alpha_{m}=\frac{\norm{\nabla_\psi F(\psi_{m})}_2^2}{ \langle \eta_{m-1},{\nabla_\psi F(\psi_{m})}-{\nabla_\psi F(\psi_{m-1})\rangle }}
  \end{aligned}
\end{equation}
It is a scalar generated from the matrices $\nabla_\psi F(\psi_{m})$, $\nabla_\psi F(\psi_{m-1})$, and $\eta_{m-1}$ through 
norm $\norm{*}$ and element-wise summation.
Such matrices to scalar conversion implies that Eq.~\eqref{eq:cg_ch1_1} demands the complete data 
of $\nabla_\psi F(\psi)$ and $\eta$, and therefore it is incompatible  with the local partial reconstruction in the basic parallelization design. 
Specifically, in the running example, the local reconstructions 
of the four GPUs generate four distinct scalars $\alpha_m$0 to $\alpha_m$3 based on the
partial $\nabla_\psi F(\psi_{m})$0 to $\nabla_\psi F(\psi_{m})$3 and $\eta_{m-1}$0 to $\eta_{m-1}$3, respectively. 
Neither of the four scalars can guarantee the convergence of the whole problem.\\

\noindent
\textit{\textbf{Challenge 2}} Line-search stage in the CG solver is a {\em while loop} that dynamical determines step length. 
It repeatedly computes the minimization function Eq.~\eqref{Eq:ML} with shrunk $\gamma_{m}$ until the condition Eq.~\eqref{Eq:linesearch} is fulfilled.
In Eq.~\eqref{Eq:ML}, $d$ and $\Gop\psi$ at the right-hand side are 3D arrays, while $F(\psi)$ at the left-hand size is 
a scalar. The array to scalar conversion is performed through the element-wise summation. Such dimensional reduction implies 
that computing $F(\psi_m+\gamma_m\eta_m)$ demands the complete data of $d$ and $\Gop\psi$. 
Therefore, similar to Eq.~\eqref{eq:cg_ch1_1}, the line-search using Eq.~\eqref{Eq:linesearch} and Eq.~\eqref{Eq:ML} is 
incompatible with the local partial reconstruction in the basic parallelization design. 
With the basic design, different GPUs will generate different 
scalar $\gamma_{m}$ and none of them can guarantee the convergence of the whole problem.\\

Addressing the above challenges requires a fine-grained parallelization with additional synchronization between GPUs for the CG solver. 
The advanced solution needs to be compatible with both Eq.~\eqref{Eq:ML} and \eqref{Eq:DaiYuan}.

\subsubsection*{Hybrid model design}\label{subsec:design2}
In our design, we propose a \textit{hybrid parallelization model} for the CG solver aiming to fully maintain its algorithmic equivalence. 
As shown in Fig.~\ref{fig:workflow}, an iteration of the CG solver can be categorized to four stages. 
In the \textit{GRAD stage}, the solver computes the gradient $\nabla_\psi F(\psi_{m})$ via Eq.~\eqref{Eq:gradient}. It then computes the search direction $\eta_{m}$ using Eq.~\eqref{Eq:DaiYuan} in the \textit{DIR stage} and the step length $\gamma_{m}$ using line-search in the \textit{LS stage}. 
It finally updates the estimated object image $\psi_{m}$ with Eq.~\eqref{Eq:update} in the \textit{Update stage}. 
The general parallelization design is applicable to the GRAD and Update stages. 
In order to address the two challenges described in the above subsection, we use the \textit{\textbf{gather-scatter}} communication pattern
for the DIR stage and the \textit{\textbf{all-reduce}} pattern for the LS stage. 
We couple the two patterns with the basic parallelization design to construct the hybrid model.
\begin{figure}[htb]
\centering
\includegraphics[width=0.75\linewidth]{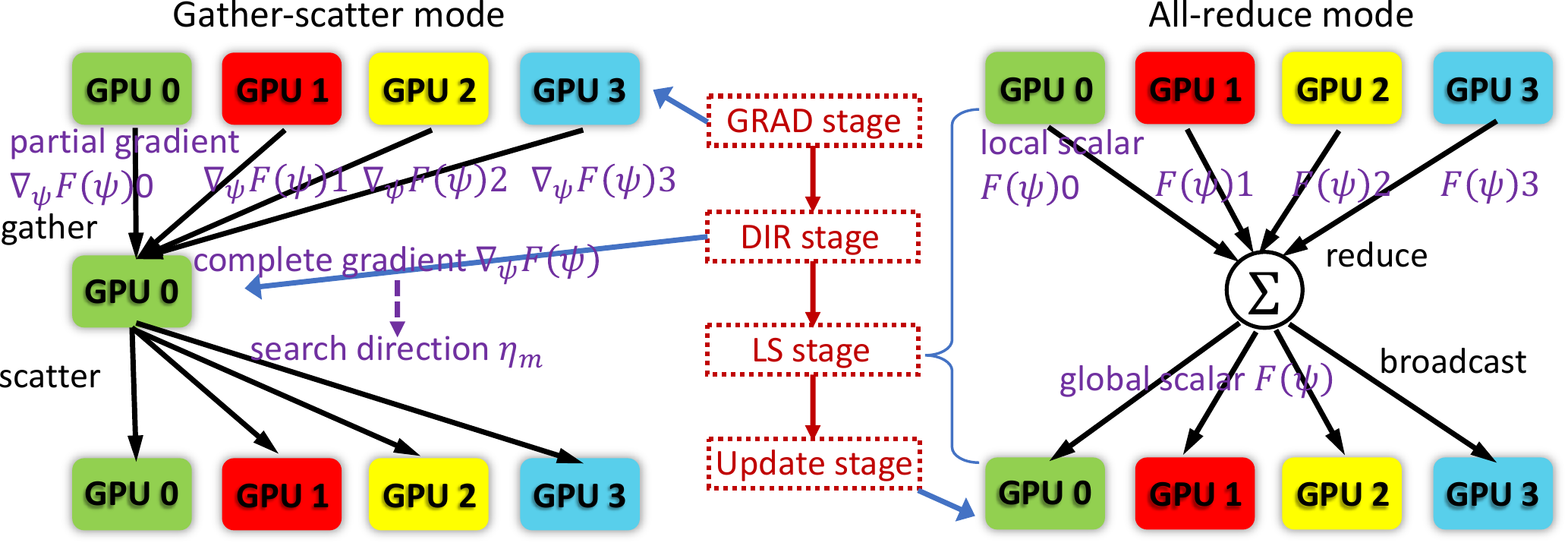}
\caption{The schematic diagram of our hybrid parallelization model. The left drawing illustrates the workflow of gather-scatter mode while the right drawing demonstrates the workflow of all-reduce mode. The gather-scatter mode is dedicated to DIR stage. The all-reduce mode is used to repeatedly compute the minimization function for LS stage.}
\label{fig:hybrid}
\end{figure}

Fig.~\ref{fig:hybrid} is a schematic depicting the hybrid parallelization model. 
The left part of Fig.~\ref{fig:hybrid} demonstrates the structure and workflow of the \textit{\textbf{gather-scatter mode}}, while the right part demonstrates the \textit{\textbf{all-reduce mode}}'s.\\

\noindent
\textit{\textbf{Gather-scatter mode:}} This mode is dedicated for the DIR stage. In such mode, one GPU is selected as 
the \textit{master GPU}. Without loss of generality, in the running example, GPU 0 is used as the \textit{master GPU}.
In the GRAD stage, the perspective partial gradients $\nabla_\psi F(\psi_{m})$0 to $\nabla_\psi F(\psi_{m})$3 
are computed and locally stored in the corresponding GPUs. 
Then in the DIR stage, all local partial gradients are gathered by the master GPU 0 to form complete 
$\nabla_\psi F(\psi_{m})$.  
The complete gradient $\nabla_\psi F(\psi_{m})$ is a 2D matrix having the same shape of the object image $\psi_{m}$.
Its data volume (usually smaller than $2048^2$) is far less than the 3D array diffraction patterns $d$ which consumes 
most of the memory. Therefore, the gathering overhead is lightweight. Moreover, since $\nabla_\psi F(\psi_{m-1})$, $\eta_{m-1}$, 
and $\nabla_\psi F(\psi_{m})$ are all 2D matrices that can fit into a single GPU (GPU 0) and Eq.~\eqref{Eq:DaiYuan} does not 
rely upon 3D data $d$ or $\Gop\psi$, the complete search direction $\eta_{m}$ can be computed entirely in master GPU 0.
Then GPU 0 splits $\eta_{m}$ as $\eta_{m}$0 to $\eta_{m}$3 and scatters them to the corresponding GPUs for the next stage.\\

\noindent
\textit{\textbf{All-reduce mode:}}
This mode is designed for the LS stage to overcome Challenge 2. As described in the previous subsection, 
Eq.~\eqref{Eq:ML} requires the complete 3D arrays $d$ and $\Gop\psi$. 
However, in contrast to the 2D array $\nabla_\psi F(\psi_{m})$, gathering the partial $d$ and $\Gop\psi$ into a master GPU is 
infeasible due to their large data volumes. 
On the other hand, we observe that the out-most element-wise summation in Eq.~\eqref{Eq:ML} converts the 3D arrays to a scalar and is the only obstacle to the local partial reconstruction. 
Therefore, according to the addition associativity, we can overcome Challenge 2 by summing up the partial results to get the global scalar $F(\psi_m+\gamma_m\eta_m)$. 
For instance, in the running example, the four GPUs compute four local scalars $F(\psi_m+\gamma_m\eta_m)$0 to $F(\psi_m+\gamma_m\eta_m)$3 via Eq.~\eqref{Eq:ML} using their perspective partial data $d$0 to $d$3, $\Gop\psi_m$0 to $\Gop\psi_m$3, $\psi_m$0 to $\psi_m$3, and $\eta_m$0 to $\eta_m$3. 
The global scalar $F(\psi_m+\gamma_m\eta_m)$ then can be calculated by $\sum_{g=0}^{3}F(\psi_m+\gamma_m\eta_m)g$ and broadcast to all GPUs. Here $\sum_{g=0}^{3}$ serves as the generalized reduction operation.\\ 

In summary, our proposed hybrid parallelization model utilizes the gather-scatter mode to overcome Challenge 1 by avoiding the computing with partial data. 
It gathers the partial data into a master GPU and performs the entire DIR stage in this GPU. 
The computed complete search direction is then divided and scattered to the corresponding GPUs. 
Our hybrid model employs the all-reduce mode to overcome Challenge 2 by reducing the partial results via a summation. 
Summing up each GPU's partial result preserves the algorithmic equivalence to Eq.~\eqref{Eq:ML}. 
The summed scalar $F(\psi_m+\gamma_m\eta_m)$ is then broadcast to all GPUs. 
The LS stage keeps iterating by shrinking the step length $\gamma_m$ and computing $F(\psi_m+\gamma_m\eta_m)$ in all-reduce mode until Eq.~\eqref{Eq:linesearch} is fulfilled. 
The hybrid model then switches back to the basic parallelization design and updates the object image $\psi_m$ accordingly.

\subsection*{PtyGer Implementation}\label{subsec:design3}
We implement the proposed hybrid parallelization model in PtyGer, which is a high-performance multi-GPU based tool for the ptychographic reconstruction.
To conform to the computational imaging domain's preference, PtyGer is written in Python using CuPy (https://cupy.chainer.org/) library. 
Alg.~\ref{alg:workflow} demonstrates the complete workflow of PtyGer. 
After partitioning and copying the inputs from the host to the GPU devices (Line~\ref{line:HtoD}), PtyGer iteratively updates the object image until the number of iteration reaches the limit (Line~\ref{line:iter}-\ref{line:itere}).
In each iteration, PtyGer first locally computes the partial  gradients in corresponding GPUs during the GRAD stage (Line~\ref{line:grads}-\ref{line:grade}), then switches to the gather-scatter mode to calculate the search-direction in the DIR stage (Line~\ref{line:dirs}-\ref{line:dire}). 
It subsequently switches to the all-reduce mode and determines the step-length via the line-search (Line~\ref{line:LSs}-\ref{line:LSe}). 
During the line-search, it repeatedly executes the minimization function (Line~\ref{line:minfs}-\ref{line:minfe}) until the stop condition is satisfied (Line~\ref{line:LSe}). 
PtyGer finally updates the subimages in every GPUs (Line~\ref{line:updates}) and exchanges their borders (Line~\ref{line:itere}). 
At the end of iterative updates, all subimages are transferred from GPUs to CPU and aligned to form the complete object image (Line~\ref{line:output}). 
PtyGer is open source and available at GitHub (https://github.com/xiaodong-yu/PtyGer). 
\begin{algorithm}[thb]	
	\SetKw{Endif}{\#endif}
	\SetKw{New}{new}
	\SetKw{Do}{do}
	\SetKw{Input}{Input:}
	\SetKw{Output}{Output:}
	\caption{Complete workflow of PtyGer}
	\label{alg:workflow}
	\LinesNumbered
	\SetNlSty{textsf}{}{}

    \Input {\fontfamily{qtm}\selectfont \textup{complex64}} $h\_\psi$, $h\_p$; {\fontfamily{qtm}\selectfont float32} $h\_s$, $h\_d$\;
    Spawn T threads, T = Number of GPUs\\
    \nonl{\footnotesize/*\textcolor{blue}{{\fontfamily{ptm}\selectfont \textit{ partition and distribute the workloads in the CPU host to different GPU devices }}}*/}\\
    \nl in thread$_i$, $i\in$ GPU\_ids: \With{\textup{GPU device $i$} \Do \Comment{{\fontfamily{ptm}\selectfont \footnotesize\textit{\textcolor{blue}{each thread is assigned a GPU device}}}}}{
        $d\_\psi_{0i}$$\leftarrow$$h\_\psi_{i}$, $d\_p_i$$\leftarrow$$h\_p_i$, $d\_s_i$$\leftarrow$$h\_s_i$, $d\_d_i$$\leftarrow$$h\_d_i$;\label{line:HtoD} \Comment{{\fontfamily{ptm}\selectfont \footnotesize\textit{\textcolor{blue}{copy data from host (h\_*) to device (d\_*)}}}}
    }
    $m=0$\;
    \While{$m$<\textup{iter\_limit}}{\label{line:iter}
        in thread$_i$, $i\in$ GPU\_ids: \Withdo{\textup{GPU device $i$}}{
            \nonl{\footnotesize/*\textcolor{blue}{{\fontfamily{ptm}\selectfont \textit{ GRAD stage }}}*/}\\
            \nl $\Gop\psi_{i}$ = FORWARD($d\_\psi_{mi}$, $d\_p_i$, $d\_s_i$)\;\label{line:grads}
            $\nabla_\psi F(\psi)_{i}$ = GRAD\_COMP($\Gop\psi_{i}$, $d\_d_i$);\label{line:grade} \Comment{{\fontfamily{ptm}\selectfont \footnotesize\textit{\textcolor{blue}{locally compute the partial gradient via Eq.(5)}}}}\\
            \nonl{\footnotesize/*\textcolor{blue}{{\fontfamily{ptm}\selectfont \textit{ DIR stage; switch to the gather-scatter mode }}}*/}\\
            \nl $\nabla_\psi F(\psi)\longleftarrow\nabla_\psi F(\psi)_{0}\cup \nabla_\psi F(\psi)_{i}$;\label{line:dirs} \Comment{{\fontfamily{ptm}\selectfont \footnotesize\textit{\textcolor{blue}{copy the partial gradient to master GPU 0 (gather)}}}}\\
        }
        thread\_sync\;
        in thread$_0$: \Withdo{\textup{GPU device $0$}}{
            $\eta$ = DAI-YUAN($\nabla_\psi F(\psi)$); \Comment{{\fontfamily{ptm}\selectfont \footnotesize\textit{\textcolor{blue}{compute the complete search direction in master GPU 0 via Eq.(7)}}}}\\
            $\eta\longrightarrow\eta_i$;\label{line:dire} \Comment{{\fontfamily{ptm}\selectfont \footnotesize\textit{\textcolor{blue}{split and distribute the search direction to slave GPUs (scatter)}}}}\\
        }
        \nonl{\footnotesize/*\textcolor{blue}{{\fontfamily{ptm}\selectfont \textit{ LS stage; switch to the all-reduce mode }}}*/}\\
        \nl $\gamma^{(0)}=1$; $\tau=0.5$; $k=0$\label{line:LSs}\;
        \Dowhile{$F(\psi+\gamma^{(k)}\eta) \leqslant F(\psi_m) + \gamma^{(k)}t$\label{line:LSe}}{
            in thread$_i$, $i\in$ GPU\_ids: \Withdo{\textup{GPU device $i$}}{
                $\gamma^{(++k)}\longleftarrow\gamma^{(k)}\tau$\;
                $\Gop(\psi+\gamma^{(k)}\eta)_{i}$ = FORWARD($d\_\psi_{mi}+\gamma^{(k)}\eta_i$, $d\_p_i$, $d\_s_i$)\label{line:minfs}\;
                $F(\psi+\gamma^{(k)}\eta)_i$ = MINFUNC($\Gop(\psi+\gamma^{(k)}\eta)_{i}$, $d\_d_i$); \Comment{{\fontfamily{ptm}\selectfont \footnotesize\textit{\textcolor{blue}{locally compute the result via Eq.(4)}}}}\\

            }
            thread\_sync\;
            $F(\psi+\gamma^{(k)}\eta)\longleftarrow\sum_{}^{}F(\psi+\gamma^{(k)}\eta)_i$;\label{line:minfe} \Comment{{\fontfamily{ptm}\selectfont \footnotesize\textit{\textcolor{blue}{sum up the partial results then broadcast it (all-reduce)}}}}\\
        }
        \nonl{\footnotesize/*\textcolor{blue}{{\fontfamily{ptm}\selectfont \textit{ Update stage; switch back to the basic design }}}*/}\\
        \nl in thread$_i$, $i\in$ GPU\_ids: \Withdo{\textup{GPU device $i$}}{
                $d\_\psi_{(++m)i}\longleftarrow d\_\psi_{mi}+\gamma^{(k)}\eta_i$\label{line:updates}\;
                exchange the border of $d\_\psi_{mi}$\label{line:itere}
        }
	}
	\Output $h\_\psi\longleftarrow\cup d\_\psi_{mi}$\label{line:output} \Comment{{\fontfamily{ptm}\selectfont \footnotesize\textit{\textcolor{blue}{copy the partial images back to host and unite them }}}}\\
\end{algorithm}


\bibliography{ref}

%

\section*{Acknowledgements (not compulsory)}

This material is based upon work supported by the U.S. Department of Energy, Office of Science, Basic Energy Sciences and Advanced Scientific Computing Research, under Contract DE-AC02-06CH11357. This research used resources of the Advanced Photon Source, a U.S. Department of Energy (DOE) Office of Science User Facility operated for the DOE Office of Science by Argonne National Laboratory under the same contract. Authors acknowledge Junjing Deng, Yudong Yao, Yi Jiang, Jeffrey Klug, Nick Sirica, and Jeff Nguyen for providing the experimental data acquired on the Velociprobe instrument at beamline 2-ID-D of the Advanced Photon Source, Argonne National Laboratory. 
This work was performed, in part, at the Center for Integrated Nanotechnologies, an Office of Science User Facility operated for the U.S. Department of Energy (DOE) Office of Science. Los Alamos National Laboratory, an affirmative action equal opportunity employer, is managed by Triad National Security, LLC for the U.S. Department of Energy’s NNSA, under contract 89233218CNA000001.
This work is partially supported by the Oﬃce of the Director of National Intelligence (ODNI), Intelligence Advanced Research Projects  Activity (IARPA), via contract D2019-1903270004.
The views and conclusions contained herein are those of the authors and should not be interpreted as necessarily representing the official policies or endorsements, either expressed or implied, of the ODNI, IARPA, or the U.S. Government.

\section*{Author contributions statement}

Must include all authors, identified by initials, for example:
A.A. conceived the experiment(s),  A.A. and B.A. conducted the experiment(s), C.A. and D.A. analysed the results.  All authors reviewed the manuscript. 

\section*{Additional information}

To include, in this order: \textbf{Accession codes} (where applicable); \textbf{Competing interests} (mandatory statement). 

The corresponding author is responsible for submitting a \href{http://www.nature.com/srep/policies/index.html#competing}{competing interests statement} on behalf of all authors of the paper. This statement must be included in the submitted article file.

\end{document}